\documentclass{ws-rv9x6}
\usepackage{subfigure}   % required only when side-by-side / subfigures are used
\usepackage{ws-rv-thm}   % comment this line when `amsthm / theorem / ntheorem` package is used
\usepackage{ws-rv-van}   % numbered citation & references (default)
\usepackage{siunitx}
\makeindex
\begin{document}

\chapter[Using World Scientific's Review Volume Document Style]{An Energy Recovery Linac for the LHeC\\\label{ra_ch1}}

\author[S.A. Bogacz, B.J. Holzer and J.A. Osborne]{S. Alex Bogacz\footnote{bogacz@jlab.org}, Bernhard J. Holzer
and John A. Osborne}
%\index[aindx]{Author, F.} % or \aindx{Author, F.}
%\index[aindx]{Author, S.} % or \aindx{Author, S.}

\address{European~Organization~for~Nuclear~Research~(CERN),\\ Gen\`eve, Switzerland \\
%f\_author@wspc.com.sg\footnote{Affiliation footnote.}}
*Center for Advanced Studies of Accelerators, Jefferson Lab (TJNAF), Newport News, USA\footnote{corresponding author}}
\begin{abstract}
The LHeC  provides an intense, high energy electron beam 
to collide with the LHC. It represents the highest energy application of energy recovery linac (ERL) technology which is increasingly recognised as one of the major pilot technologies for the development of particle physics because it utilises and stimulates superconducting RF technology progress, and it
 increases intensity while keeping the power consumption low. The LHeC instantaneous luminosity is determined through the integrated luminosity goal. The electron beam energy is chosen to achieve  TeV cms collision energy and enable competitive searches 
and precision Higgs boson measurements. The wall-plug power has been constrained to $100$\,MW. Two super-conducting linacs of about $900$\,m length, which are placed opposite to each other,
accelerate the passing electrons by $8.3$\,GeV each. This
leads to a final electron beam energy of about $50$\,GeV in 
a 3-turn racetrack energy recovery linac  configuration.
 \end{abstract}

\body

%\tableofcontents

\section{The ERL Configuration of the LHeC}

The LHeC  provides an intense, high energy electron beam 
to collide with the LHC.  It represents the highest energy application of energy recovery linac (ERL) technology which is increasingly recognised as one of the major pilot technologies for the development of particle physics because
 it utilises and stimulates superconducting RF technology progress, and it increases intensity while keeping the power consumption low.
The electron beam energy is chosen to achieve  TeV cms collision energy  and enable competitive searches 
and precision Higgs boson measurements. A cost-physics-energy evaluation is presented here which points to choosing $E_e \simeq 50$\,GeV as a new 
default value, which was $60$\,GeV before.
The wall-plug power has been constrained to $100$\,MW. 
Two super-conducting linacs of about $900$\,m length, which are placed opposite to each other,
accelerate the passing electrons by $8.3$\,GeV each. This
leads to a final electron beam energy of about $50$\,GeV in 
a 3-turn racetrack energy recovery linac  configuration.
Cost considerations and machine--detector performance aspects, in particular the amount of synchrotron radiation losses in the IR, have led to define a new reference configuration with $E_e = \SI{49.2}{GeV}$ and a circumference of $\approx\SI{5.4}{km}$, 1/5 of that of the LHC.

\begin{figure}[tbh]
  \centering
  \includegraphics[width=0.8\textwidth]{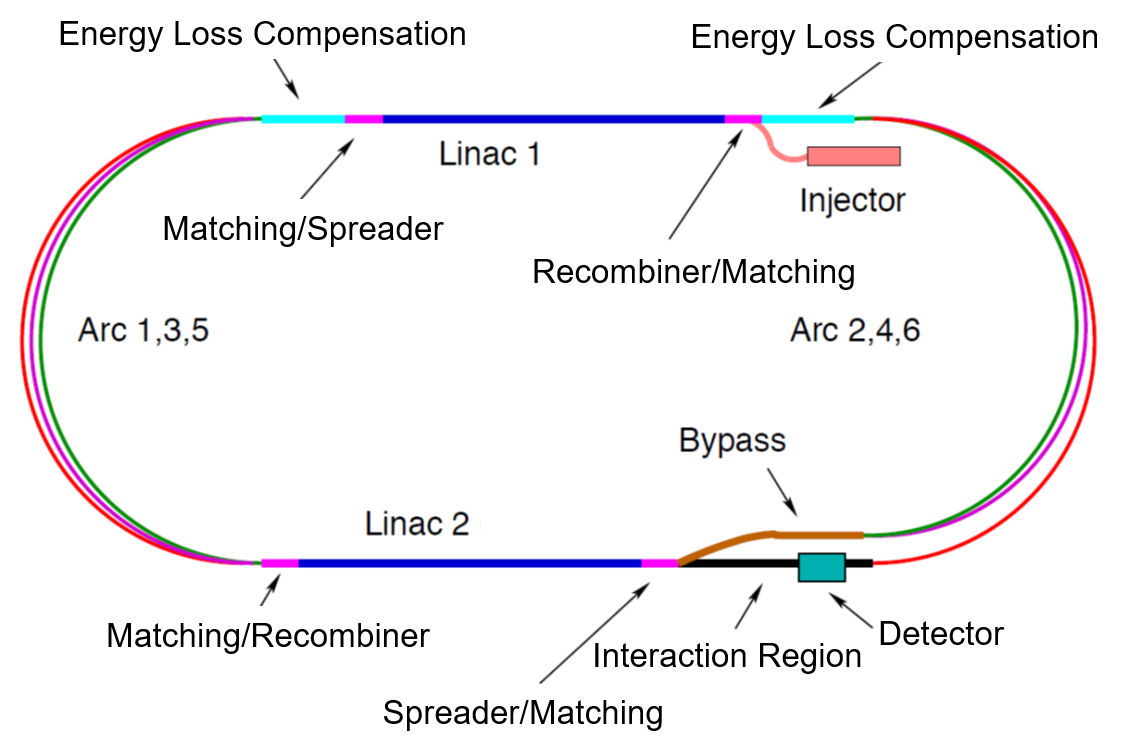}
  \caption{Schematic layout of the LHeC design based on an Energy Recovery Linac.}
  \label{fig:ERL_sketch}
\end{figure}

The ERL consists of two superconducting (SC) linacs operated in CW connected by at least three pairs of arcs to allow three accelerating and three decelerating passes (see Fig.~\ref{fig:ERL_sketch}). The length of the high energy return arc following the interaction point should be such as to provide a half RF period wavelength shift to allow the deceleration of the beam in the linac structures in three passes down to the injection energy and its safe disposal.  SC Cavities with an unloaded quality factor $Q_0$ exceeding $10^{10}$ are required to minimise the requirements on the cryogenic cooling power and to allow an efficient ERL operation. The choice of having three accelerating and three decelerating passes implies that the circulating current in the linacs is six times the current colliding at the Interaction Point (IP) with the hadron beam.

The main parameters of the LHeC ERL are listed in Tab.~\ref{tab:ERLparameters}; their choices and optimisation criteria will be discussed in the following sections.
\begin{table}[!ht]
  \centering
  \small
 \begin{tabular}{lcc} 
    \toprule
    Parameter & Unit & Value  \\
    \botrule  
    Injector energy & \si{GeV} & 0.5 \\
    Total number of linacs & & 2 \\
    Number of acceleration passes & & 3 \\
    Maximum electron energy & \si{GeV} & 49.19 \\
    Bunch charge & \si{pC} & $499$ \\
    Bunch spacing & \si{ns}	& 24.95 \\
    Electron current & \si{mA}	& 20 \\
    Transverse normalized emittance & \si{\micro\meter} & 30 \\
    Total energy gain per linac & \si{GeV} & 8.114\\
    Frequency & \si{MHz} & 801.58  \\
    Acceleration gradient & \si{MV/m} & 19.73 \\
    Cavity iris diameter & \si{\milli\meter} & 130 \\
    Number of cells per cavity & & 5 \\
    Cavity length (active/real estate) & \si{\meter} & 0.918/1.5 \\
    Cavities per cryomodule &  & 4  \\
    Cryomodule length & \si{m} & 7 \\
    Length of 4-CM unit & \si{m} & 29.6 \\
    Acceleration per cryomodule (4-CM unit) & \si{MeV} & 289.8 \\
    Total number of cryomodules (4-CM units) per linac & & 112 (28) \\
    Total linac length (with with spr/rec matching) & \si{m} & 828.8 (980.8) \\
    Return arc radius (length) & \si{m} & 536.4 (1685.1) \\
    Total ERL length & \si{km} & 5.332 \\
\botrule
 \end{tabular}
 \caption{Parameters of LHeC Energy Recovery Linac (ERL).}
 \label{tab:ERLparameters}
\end{table}
\\
\subsection{Baseline Design -- Lattice Architecture  }
The ERL, as sketched in Fig.~\ref{fig:ERL_sketch}, is arranged in a racetrack configuration; hosting two superconducting linacs in the parallel straights and three recirculating arcs on each side. The linacs are \SI{828.8}{m} long and the arcs have \SI{536.4}{m} radius, additional space of \SI{76}{m} is taken up by utilities like Spreader/Recombiner (Spr/Rec), matching and energy loss compensating sections adjacent to both ends of each linac (total of 4 sections)~\cite{Pellegrini:2015rdx}. The total length of the racetrack is \SI{5.332}{km}: 1/5 of the LHC circumference $2 \cdot (828.8+2\cdot 76 +  536.4\cdot\pi)~\si{m}$. Each of the two linacs provides 8.114 GV accelerating voltage, therefore a \SI{49.19}{GeV} energy is achieved in three turns. After the collision with the protons in the LHC, the beam is decelerated in the three subsequent turns. The injection and dump energy has been chosen at \SI{0.5}{GeV}.

Injection into the first linac is done through a fixed field injection chicane, with its last magnet (closing the chicane) being placed at the beginning of the linac.
It closes the orbit \emph{bump} at the lowest energy, injection pass, but the magnet (physically located in the linac) will deflect the beam on all subsequent linac passes. In order to close the resulting higher pass \emph{bumps}, the so-called re-injection chicane is instrumented, by placing two additional opposing bends in front of the last chicane magnet.
The chosen arrangement is such that, the re-injection chicane magnets are only \emph{visible} by the higher pass beams.
The second linac in the racetrack is configured exactly as a mirror image of the first one, with a replica of the re-injection chicane at its end, which facilitates a fixed-field extraction of energy recovered beam to the dump.

\begin{figure}[t]
  \centering
  \includegraphics[width=0.95\textwidth]{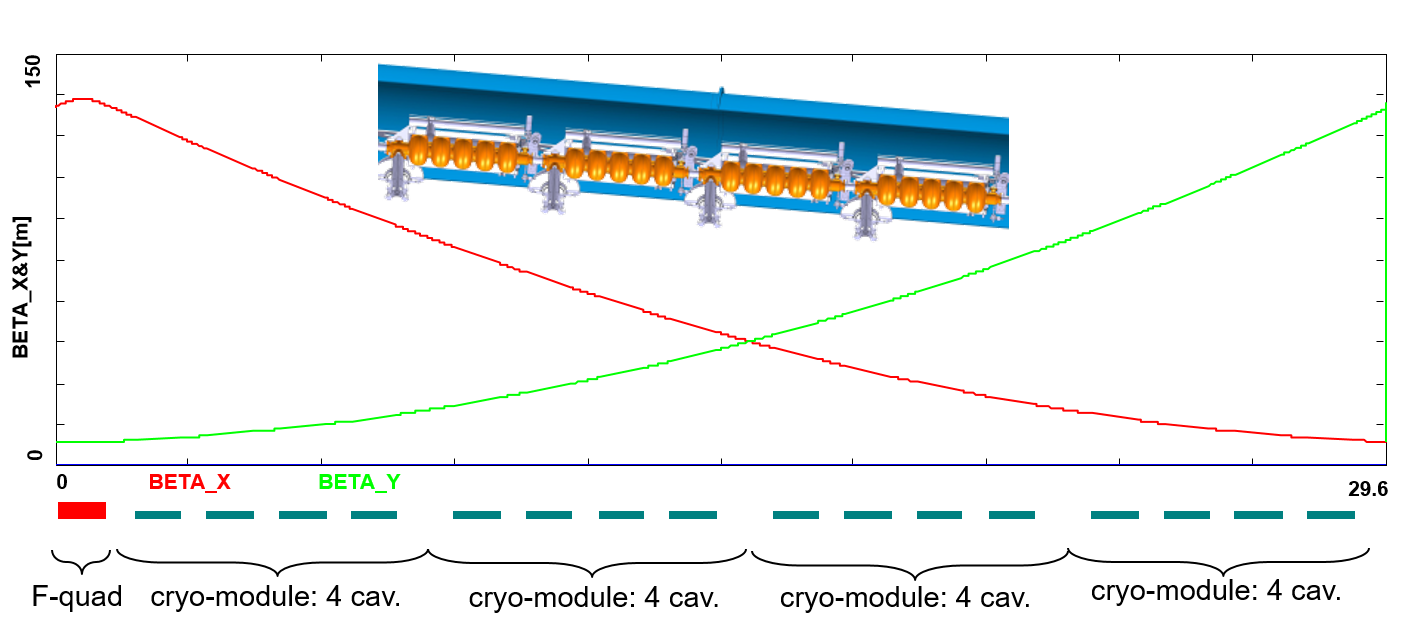}
  \caption{Layout of a half-cell composed out of four cryomodules (each hosting four, 5-cell cavities: top insert) and a focusing quad. Beta functions reflect $130^0$ FODO optics.}
  \label{fig:Half_cell}
\end{figure}

\subsubsection{Linac Configuration and Multi-pass Optics}
Appropriate choice of the linac optics is of paramount importance for the transverse beam dynamics in a multi-pass ERL. The focusing profile along the linac (quadrupole gradients) need to be set (and they stay constant), so that multiple pass beams within a vast energy range may be transported efficiently. The chosen arrangement is such that adequate transverse focusing is provided for a given linac aperture. The linac optics is configured as a strongly focusing, $130^0$ FODO. In a basic FODO cell a quadrupole is placed every four cryomodules, so that the full cell contains two groups of 16 RF cavities and a pair of quads (F, D) as illustrated in Fig.~\ref{fig:Half_cell}. The entire linac is built out of 14 such cells.
Energy recovery in a racetrack topology explicitly requires that both the
accelerating and decelerating beams share the individual return arcs~\cite{ICFA_Bogacz}. This in turn, imposes specific requirements for TWISS function at the linacs ends: TWISS functions have to be identical for both the accelerating and decelerating linac passes converging to the same energy and therefore entering the same arc. There is an alternative scheme, proposed by Peter Williams~\cite{erl.15}, who has argued that it would be beneficial to separate the accelerating and decelerating arcs. This would simplify energy compensation systems and linac-to-arc matching, but at an higher cost of the magnetic system of the arcs. However, doubling number of arcs is a very costly proposition. On the other hand, C-BETA experiment is pioneering a multi-pass arcs to transport a vast energy range through the same beam-line and it still intends to use them for energy recovery. Our approach, based on proven, CEBAF-like, RLA technology~\cite{CEBAF_12} is somewhere in the 'middle'. 

\begin{figure}[th]
  \centering
  \includegraphics[width=1.0\textwidth]{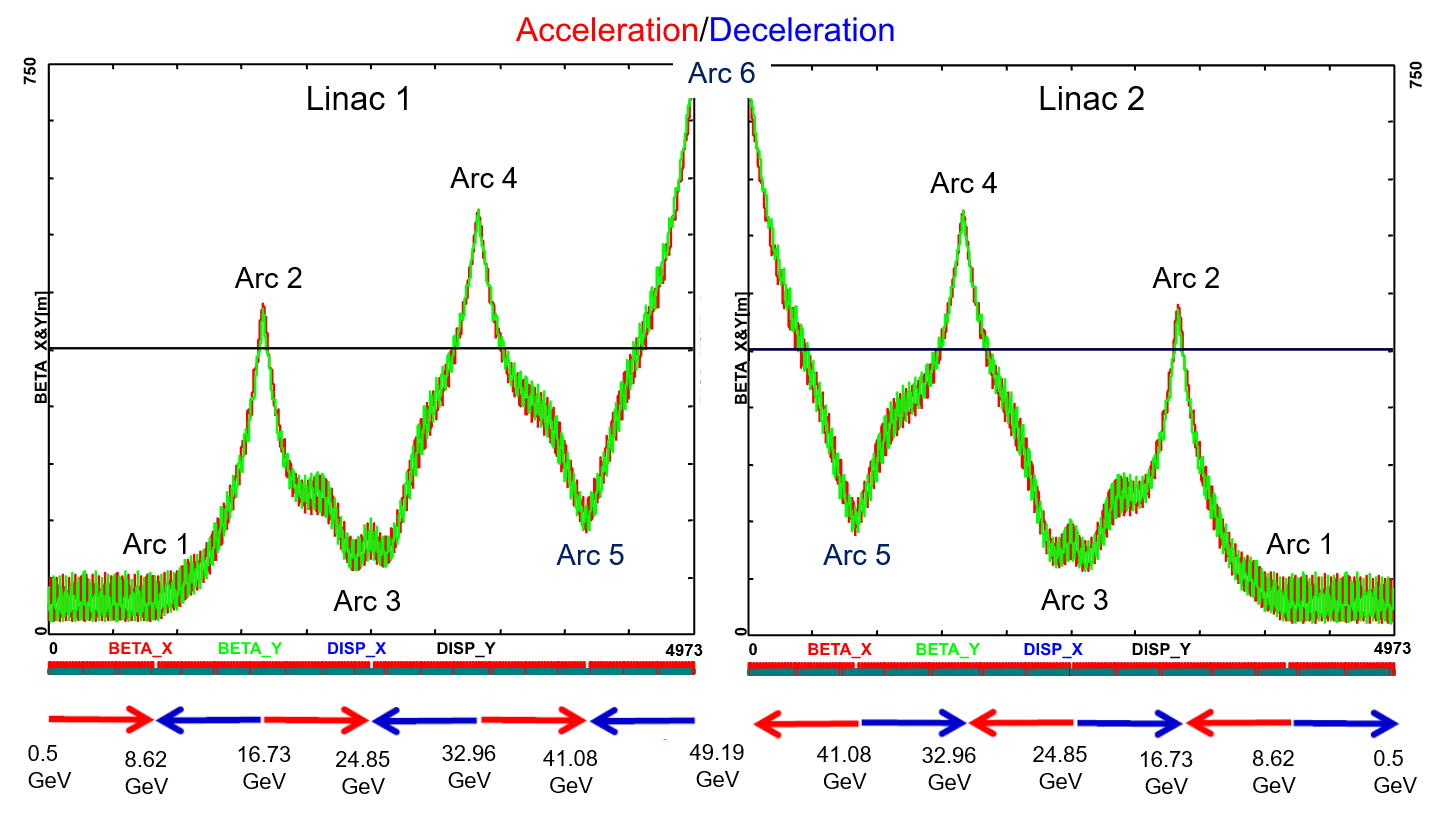}
  \caption{Beta function in the optimised multi-pass linacs (3 accelerating passes and 3 decelerating passes in each of two linacs). The matching conditions are automatically built into the resulting multi-pass linac beamline.}
  \label{fig:Multi_pass}
\end{figure}

To visualize beta functions for multiple accelerating and decelerating passes through a given linac, it is convenient to reverse the linac direction for all decelerating passes and string them together with the interleaved accelerating passes, as illustrated in Fig.~\ref{fig:Multi_pass}.
This way, the corresponding accelerating and decelerating passes are joined together at the arc's entrance/exit.
Therefore, the matching conditions are automatically built into the resulting multi-pass linac beamline. One can see that both linacs uniquely define the TWISS functions for the arcs: Linac 1 fixes input to all odd arcs and output to all even arcs, while Linac 2 fixes input to all even arcs and output to all odd arcs.
The optics of the two linacs are mirror-symmetric; They were optimised so that, Linac 1 is periodic for the first accelerating pass  and Linac 2 has this feature for  last decelerating one.
In order to maximize the BBU threshold current~\cite{Hoffstaetter:2004qy}, the optics is tuned so that the integral of $\beta/E$ along the linac is minimised. The resulting phase advance per cell is close to $130^0$.
Non-linear strength profiles and more refined merit functions were tested, but they only brought negligible improvements.

\subsubsection{Recirculating Arcs -- Emittance Preserving Optics}
Synchrotron radiation effects on beam dynamics, such as the transverse emittance dilution induced by quantum excitations have a paramount impact on the collider luminosity. All six horizontal arcs are accommodated in a tunnel of \SI{536.4}{m} radius.  
The transverse emittance dilution accrued through a given arc is proportional to the emittance dispersion function, $H$, averaged over all arc's bends~\cite{Schwinger:1996mc}: % as expressed by Eq.~\eqref{eq:Emit_dil_0}:
\begin{equation}
  \Delta \epsilon = \frac{2 \pi}{3} C_q r_0 <H> \frac{\gamma^5}{\rho^2}\,,
  \label{eq:Emit_dil_0}
\end{equation}
where
\begin{equation}
  C_q = \frac{55}{32 \sqrt{3}} \frac{\hbar}{m c}
  \label{eq:C_q}
\end{equation}
and $r_0$ is the classical electron radius and $\gamma$ is the Lorentz boost.
Here,  $H = (1+\alpha^2)/\beta \cdot D^2 + 2 \alpha \ D D' + \beta \cdot D'^2$ where $D, D'$ are the bending plane dispersion and its derivative, with $<...>~=~\frac{1}{\pi}\int_\text{bends}...~\text{d}\theta$.

Therefore, emittance dilution can be mitigated through appropriate choice of arc optics (values of $\alpha, \beta, D, D'$ at the bends). In the presented design, the arcs are configured with a FMC (Flexible Momentum Compaction) optics to ease individual adjustment of, $<H>$, in various energy arcs. 

\begin{figure}[th]
  \centering
  \includegraphics[width=1.0\textwidth]{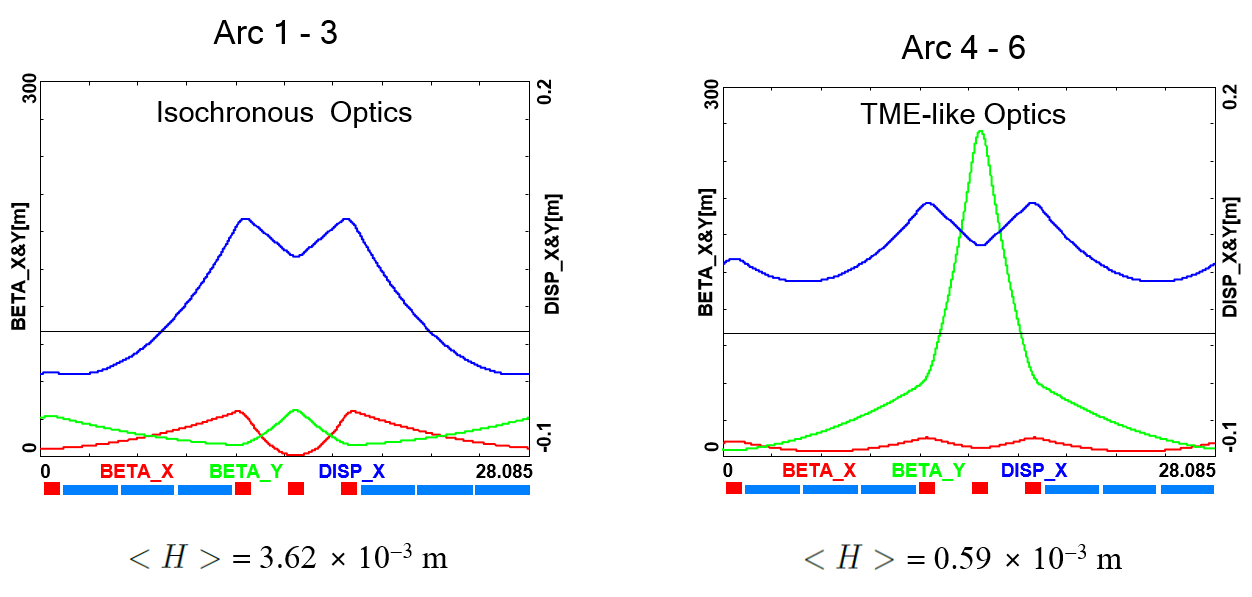}
  \caption{Two styles of FMC cells appropriate for different energy ranges. Left: lower energy arcs (Arc 1--3) configured with \emph{Isochronous} cells, Right: higher energy arcs configured with \emph{TME-like} cells. Corresponding values of the emittance dispersion averages, $<H>$, are listed for both style cells.}
  \label{fig:Arc_cells}
\end{figure}

Optics design of each arc takes into account the impact of synchrotron radiation at different energies. At the highest energy, it is crucial to minimise the emittance dilution due to quantum excitations; therefore, the cells are tuned to minimise the emittance dispersion, $H$, in the bending sections, as in the TME (Theoretical Minimum Emittance) lattice. On the other hand, at the lowest energy, it is beneficial to compensate for the bunch elongation with isochronous optics. The higher energy arcs (4, 5 and 6) configured with the TME cells are still quasi-isochronous. To fully compensate remnant bunch elongation one could set higher pass linacs slightly off-crest to compress the bunches, since one has full control of gang-phases for individual linac passes. 
All styles of FMC lattice cells, as illustrated in Fig.~\ref{fig:Arc_cells}, share the same footprint for each arc. This allows us to stack magnets on top of each other or to combine them in a single design. Here, we use substantially shorter then in the 60 GeV design, \SI{28.1}~~{m}, FMC cell configured with six \SI{3}~~{m} bends, in groups of flanked by a quadrupole singlet and a triplet, as illustrated in Fig.~\ref{fig:Arc_cells}.
The dipole filling factor of each cell is \SI{63}\%; therefore, the effective bending radius $\rho$ is \SI{336.1}~~{m}.
Each arc is followed by a matching section and a recombiner (mirror symmetric to spreader and matching section). Since the linacs are mirror-symmetric, the matching conditions described in the previous section, impose mirror-symmetric arc optics (identical betas and sign reversed alphas at the arc ends).

Path-length adjusting chicanes were also foreseen to tune the beam time of flight in order to hit the proper phase at each linac injection. Later investigations proved them to be effective only with lower energy beams, as these chicanes trigger unbearable energy losses, if applied to the highest energy beams. A possible solution may consist in distributing the perturbation along the whole arc with small orbit excitation. This issue will be fully addressed in a subsequent section on 'Synchrotron Radiation Effects - Emittance Dilution'.

% ---
\subsubsection{Spreaders and Recombiners}
\begin{figure}
  \centering
  \includegraphics[width=0.95\textwidth]{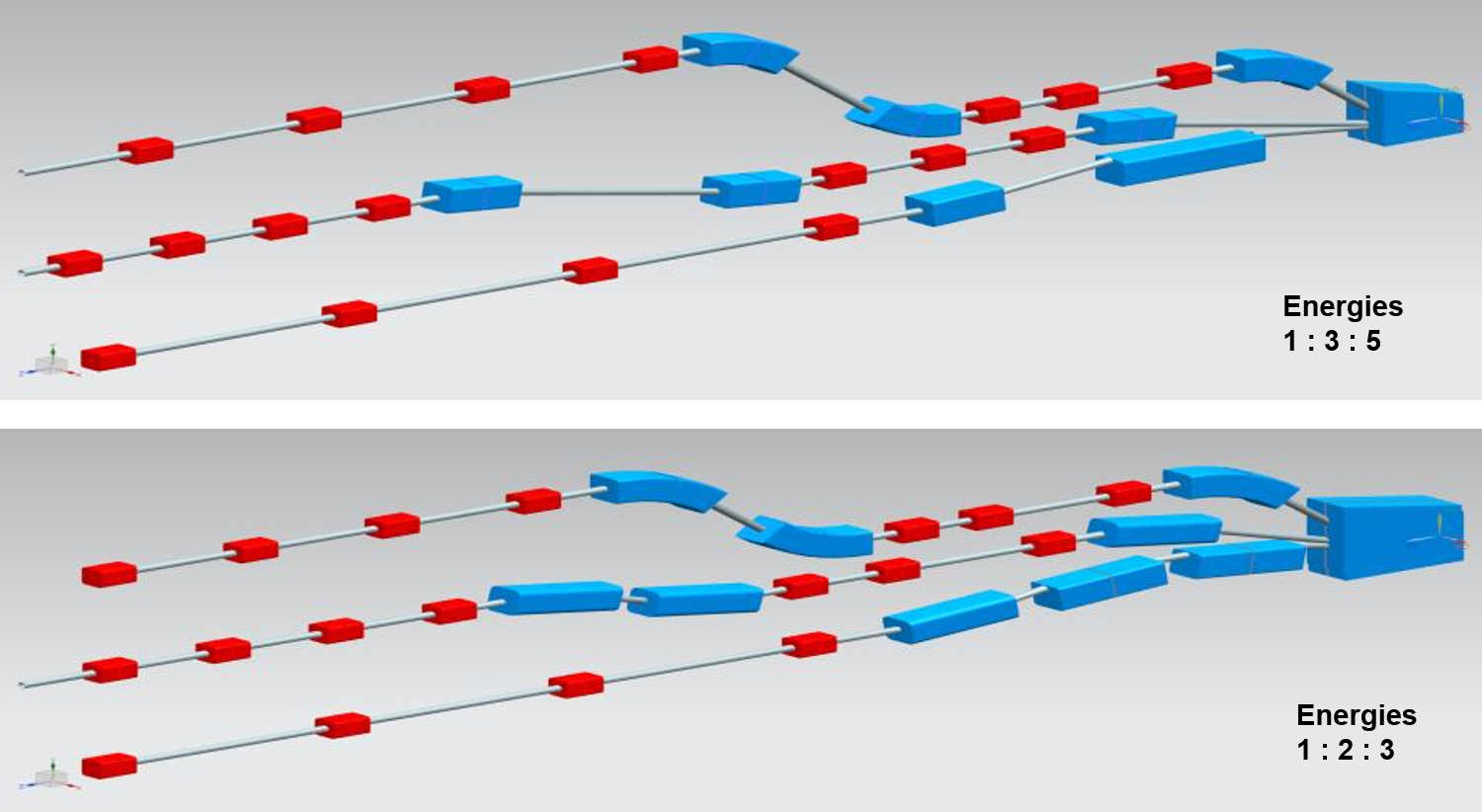}
  \caption{Layout of a three-beam switch-yard for different energy ratios: 1\,:\,3\,:\,5 and  1\,:\,2\,:\,3 corresponding to specific switch-yard geometries implemented on both sides of the racetrack}
  \label{fig:Switchyard}
\end{figure}

% ---
The spreaders are placed directly after each linac to separate beams of different energies and to route them to the corresponding arcs. The recombiners facilitate just the opposite: merging the beams of different energies into the same trajectory before entering the next linac.
As illustrated in Fig.~\ref{fig:Switchyard}, each spreader starts with a vertical bending magnet, common for all three beams, that initiates the separation. The highest energy, at the bottom, is brought back to the horizontal plane with a chicane.
The lower energies are captured with a two-step vertical bending adapted from the CEBAF design~\cite{CEBAF_12}.

\begin{figure}
  \centering
  \includegraphics[width=0.6\textwidth]{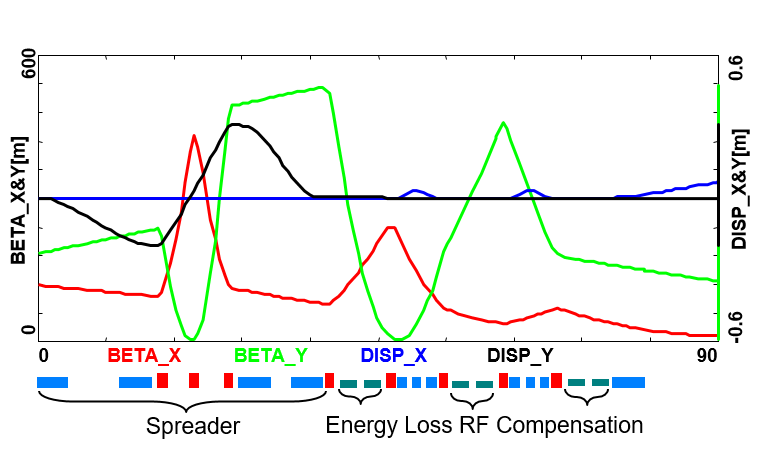}
  \caption{Spreader 3 (\SI{24.8}{GeV}) optics; featuring a vertical achromat with three dispersion suppressing quads in-between the two steps, a pair of path-length adjusting dogleg chicanes and four betatron matching quads, interleaved with three energy loss compensating sections (2-nd harmonic RF cavities marked in green).}
  \label{fig:Spreader}
\end{figure}

\begin{figure}
  \centering
  \includegraphics[width=1.0\linewidth]{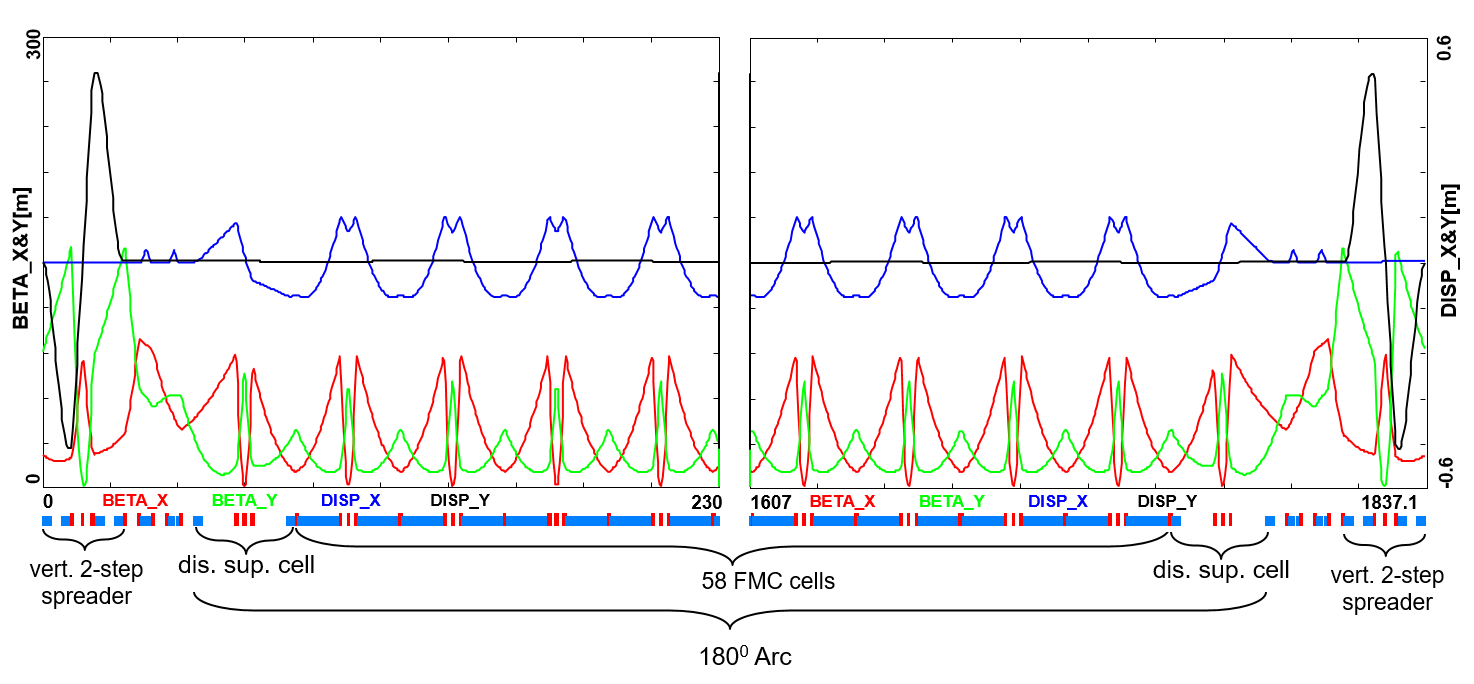}
  \caption{Complete Optics for Arc~3 (including switch-yard); featuring: low emittance \SI{180}{\degree} arc based on isochronous cells (30 cells flanked by dispersion suppression cell with missing dipoles on each side), spreaders and recombiners with matching sections and doglegs symmetrically placed on each side of the arc proper.}
  \label{fig:Arc}
\end{figure}

Functional modularity of the lattice requires spreaders and recombiners to be achromats (both in the horizontal and vertical plane). To facilitate that, the vertical dispersion is suppressed by a pair of quadrupoles located in-between vertical steps; they naturally introduce strong vertical focusing, which needs to be compensated by the middle horizontally focusing quad. The overall spreader optics is illustrated in Fig.~\ref{fig:Spreader}. 
Complete layout of two styles of switch-yard with different energy ratios is depicted in Fig.~\ref{fig:Switchyard}. Following the spreader, there are four matching quads to bridge the Twiss function between the spreader and the following $180^0$ arc (two betas and two alphas). 
Combined spreader-arc-recombiner optics, features a high degree of modular functionality to facilitate momentum compaction management, as well as orthogonal tunability for both the beta functions and dispersion, as illustrated in Fig.~\ref{fig:Arc}. 

\begin{figure}
  \centering
  \includegraphics[width=1.0\linewidth]{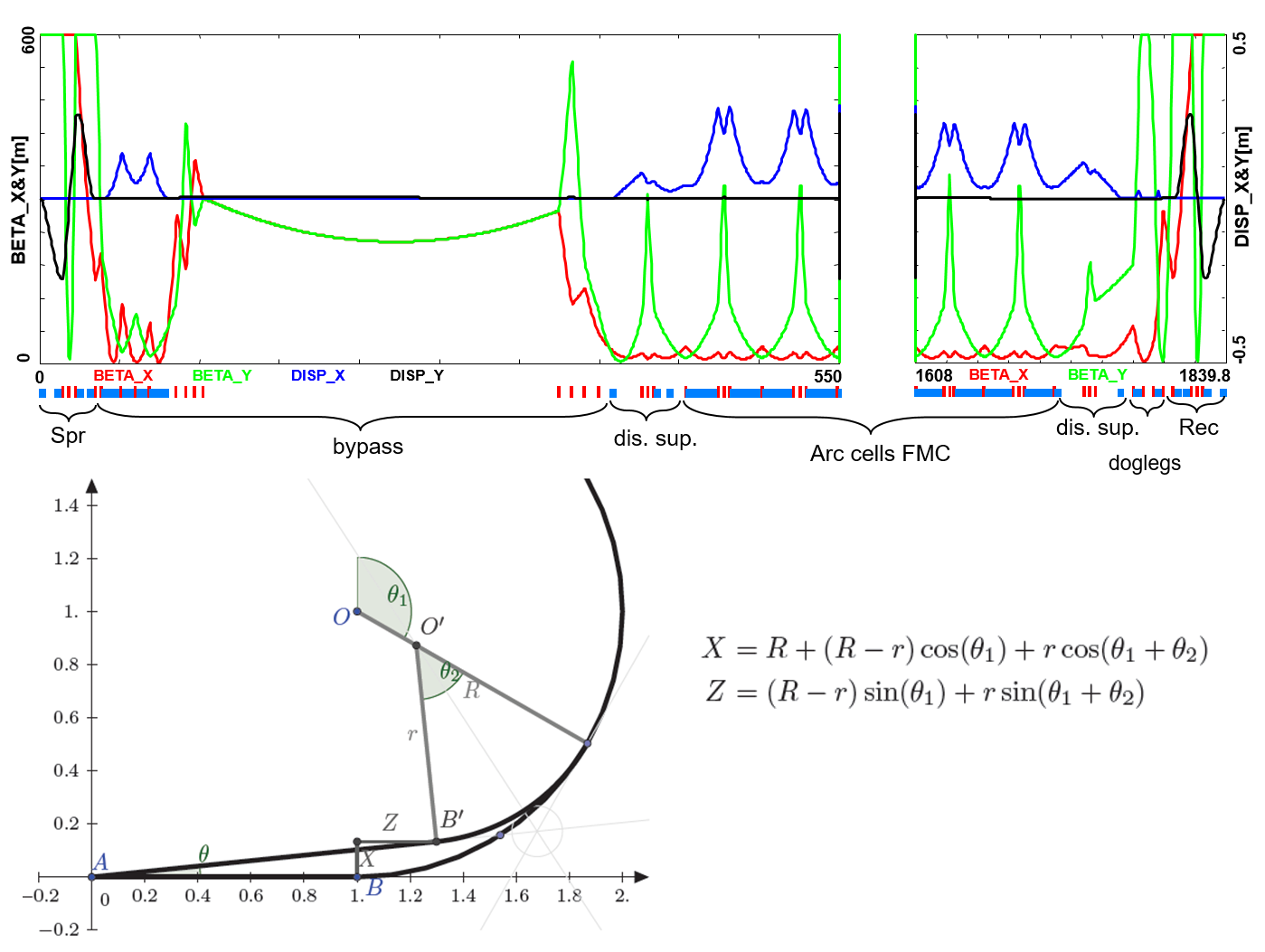}
  \caption{Optics and layout of Arc~4 including the detector bypass. The lattice (top insert) features a vertical spreader, an initial horizontal bending, a straight section, a modified dispersion suppressor, seven junction cells, and four regular cells. The bypass geometry (bottom insert), features a long IP line, AB,  which for visual reasons has been purposely stretched, being actually about $1/5$ of the arc radius. All geometric dependencies of the bypass parameters are summarized in the inserted formulae.
  }
  \label{fig:Bypass}
\end{figure}
% ---
% ---
\subsubsection {IR Bypasses}
% ---
After the last spreader the \SI{49.19}~~{GeV} beam goes straight to the interaction region. However the lower energy beams; at \SI{16.7} and \SI{33.0}~~{GeV}, need to be further separated horizontally in order to avoid interference with the detector. Different design options for the bypass section were explored~\cite{Thesis_Pellegrini} and the one that minimises the extra bending has been chosen and implemented in the lattice.

Ten arc-like dipoles are placed very close to the spreader, to provide an initial bending, $\theta$, which results in $X=\SI{10}~~{m}$ separation from the detector located \SI{120}~~{m} downstream. The straight section of the bypass is approximately \SI{240}~~{m} long. After the bypass, in order to reconnect to the footprint of Arc~6, 7 of 30 standard cells in Arc~2 and Arc~4 are replaced with 7 higher field, junction cells. The number of junction cells is a compromise between the field strength increase and the length of additional bypass tunnel, as can be inferred from the scheme summarised in Fig.~\ref{fig:Bypass}.
The stronger bending in the junction cells creates a small mismatch, which is corrected by adjusting the strengths of the quadrupoles in the last junction cell and in the first regular cell.
\subsubsection{Synchrotron Radiation Effects -- Emittance Dilution}
% ---
ERL efficiency as a source of multi-GeV electrons for a high luminosity collider is limited by the incoherent synchrotron radiation effects on beam dynamics; namely the transverse emittance dilution and the longitudinal momentum spread (induced by quantum excitations).
The first effect, the transverse emittance increase, will have a paramount impact on the collider luminosity, due to stringent limits on the allowed emittance increase.
The second one, accrued momentum spread, governs asymmetries of accelerated and decelerated beam profiles. These asymmetries substantially complicate multi-pass energy recovery and matching, and ultimately they limit the energy reach of the ERLs due to recirculating arc momentum acceptance.

Arc optics was designed to ease individual adjustment of momentum compaction (needed for the longitudinal phase-space control, essential for operation with energy recovery) and the horizontal emittance  dispersion, $H$, in each arc.
Tab.~\ref{tab:SREmittance} lists arc-by-arc dilution of the transverse, $\Delta \epsilon$, and longitudinal, $\Delta \sigma_{\frac{\Delta E}{E}}$, emittance due to quantum excitations calculated using analytic formulas, Eqs.~\eqref{eq:Emit_dil_1},~\eqref{eq:Emit_dil_2} and~\eqref{eq:Emit_dil_3}, introduced by M.~Sands~\cite{Schwinger:1996mc}:  
\begin{equation}
  \Delta E = \frac{2 \pi}{3} r_0 ~mc^2~ \frac{\gamma^4}{\rho}\,
  \label{eq:Emit_dil_1}
\end{equation}
\begin{equation}
  \Delta \epsilon_N = \frac{2 \pi}{3} C_q r_0 <H> \frac{\gamma^6}{\rho^2}\,,
  \label{eq:Emit_dil_2}
\end{equation}
\begin{equation}
  \frac{\Delta \epsilon_E^2}{E^2} = \frac{2 \pi}{3} C_q r_0~ \frac{\gamma^5}{\rho^2}\,,
  \label{eq:Emit_dil_3}
\end{equation}
where $C_q$ is given by Eq.~\eqref{eq:C_q}.
Here, $\Delta \epsilon^2_E$ is an increment of energy square variance, $r_0$ is the classical electron radius, $\gamma$ is the Lorentz boost and $C_q \approx 3.832 \cdot 10^{-13}\,\text{m}$ for electrons (or positrons).

\begin{table}[!ht]
  \centering
%  \small
  \begin{tabular}{lcccc} 
  \toprule
%  Beamline & Beam energy [\si{GeV}] & $\Delta E$ [\si{MeV}]& $\Delta \epsilon^x_N$ [\si{mm~mrad}] & $\Delta \sigma_{\frac{\Delta E}{E}}$ [\si{\percent}]\\
  Beamline & Beam energy & $\Delta E$  & $\Delta \epsilon^x_N$ & $\Delta \sigma_{\frac{\Delta E}{E}}$  \\
    & [\si{GeV}] &  [\si{MeV}]&  [\si{mm~mrad}] &  [\%]\\
    \botrule  
  arc 1 & 8.62 & 0.7 & 0.0016 & 0.0005\\
  arc 2 & 16.73 & 10 & 0.085 & 0.0027\\
  arc 3 & 24.85 & 49 & 0.91 & 0.0072\\
  arc 4 & 32.96 & 152 & 0.81 & 0.015\\
  arc 5 & 41.08 & 368 & 3.03 & 0.026\\
  arc 6 & 49.19 & 758 & 8.93 & 0.040\\
\botrule
  \end{tabular}
  \caption{Energy loss and emittance dilution (horizontal and longitudinal) due to synchroton radiation generated by all six \SI{180}{\degree} arcs (not including Spreaders, Recombiners and Doglegs). Here, $\Delta \sigma_{\frac{\Delta E}{E}} = \sqrt{\tfrac{\Delta \epsilon_E^2}{E^2}}$}.
  \label{tab:SREmittance}
\end{table}

\begin{table}[!ht]
  \centering
%  \small
  \begin{tabular}{lcccc} 
    \toprule
%    Beamline & Beam energy [\si{GeV}] & $\Delta E$ [\si{MeV}]& $\Delta \epsilon^y_N$ [\si{mm~mrad}] & $\Delta \sigma_{\frac{\Delta E}{E}}$ [\si{\percent}]\\
    Beamline & Beam energy & $\Delta E$ & $\Delta \epsilon^y_N$& $\Delta \sigma_{\frac{\Delta E}{E}}$\\
     &  [\si{GeV}] & [\si{MeV}]&  [\si{mm~mrad}] & [\%]\\
    \botrule  
    Spr/Rec 1 & 8.62 & 0.2 & 0.035 & 0.0008\\
    Spr/Rec 2 & 16.73 & 3.0 & 0.540 & 0.0044\\
    Spr/Rec 3 & 24.85 & 6.0 & 0.871 & 0.0066\\
    Spr/Rec 4 & 32.96 & 21.6 & 5.549 & 0.0143\\
    Spr/Rec 5 & 41.08 & 7.1 & 0.402 & 0.0062\\
    Spr/Rec 6 & 49.19 & 39.2 & 3.92 & 0.0205\\
\botrule
  \end{tabular}
  \caption{Energy loss and  emittance dilution (vertical and longitudinal) due to synchroton radiation generated by the two step Spreader, or Recombiner design of a given arc. Here, $\Delta \sigma_{\frac{\Delta E}{E}} = \sqrt{\frac{\Delta \epsilon_E^2}{E^2}}$}.
  \label{tab:SprRecEmittance}
\end{table}

Similarly, the horizontal emittance dilution induced by the Doglegs (four dogleg chicanes per arc) in various arcs is summarized in Tab.~\ref{tab:DogEmittance}. Each dogleg chicane is configured with four 1 meter bends (1 Tesla each), so that they bend the lowest energy beam at \SI{8.6}~~{GeV} by 2 degrees. The corresponding path-lengths gained in the Doglegs of different arcs are also indicated.

\begin{table}[!ht]
  \centering
%  \small
  \begin{tabular}{lccccc} 
    \toprule
%    Beamline & Beam energy [\si{GeV}] & $\Delta E$ [\si{MeV}]& $\Delta \epsilon^y_N$ [\si{mm~mrad}] & $\Delta \sigma_{\frac{\Delta E}{E}}$ [\%] & path-length [\si{mm}] \\
    Beamline & Beam energy  & $\Delta E$ & $\Delta \epsilon^x_N$  & $\Delta \sigma_{\frac{\Delta E}{E}}$  & path-length\\
     & [\si{GeV}] & [\si{MeV}]& [\si{mm~mrad}] & [\%] & [\si{mm}] \\
    \botrule  
    Doglegs 1 & 8.62 & 2 & 0.201 & 0.007 & 7.32\\
    Doglegs 2 & 16.73 & 9 & 0.667 & 0.009 & 1.96\\
    Doglegs 3 & 24.85 & 19 & 5.476 & 0.014 & 0.84\\
    Doglegs 4 & 32.96 & 33 & 5.067 & 0.014 & 0.52\\
    Doglegs 5 & 41.08 & 52 & 12.067 & 0.028 & 0.36\\
    Doglegs 6 & 49.19 & 74 & 2.836 & 0.011 & 0.28\\
\botrule
\end{tabular}
  \caption{Energy loss and  emittance dilution (horizontal and longitudinal) due to synchroton radiation generated by the Doglegs (four dogleg chicanes) of a given arc. Here, $\Delta \sigma_{\frac{\Delta E}{E}} = \sqrt{\frac{\Delta \epsilon_E^2}{E^2}}$}.
  \label{tab:DogEmittance}
\end{table}

As indicated in Tab.~\ref{tab:DogEmittance}, the Doglegs in the highest energy arcs, Arc 5 and Arc 6, provide only sub mm path-length gain with large synchrotron radiation effects. They are not very effective and generate strong, undesired emittance dilution. Therefore, it is reasonable to eliminate them from both Arc 5 and 6. Instead,  one could resort to an alternative path-length control via appropriate orbit steering with both horizontal and vertical correctors present at every girder and distributed evenly throughout the arc.
Combining all three contributions: ($180^0$ arc, Spreader, Recombiner and Doglegs (no Doglegs in Arcs 5 and 6), the net cumulative emittance dilution is summarized in Tab.~\ref{tab:CummEmittance} for the case of the two-step spreader.

\begin{table}[!ht]
  \centering
%  \small
  \begin{tabular}{lccccc} 
    \toprule
%    Beamline & Beam energy [\si{GeV}] & $\Delta E$ [\si{MeV}]& $\Delta^{cum} \epsilon^x_N$ [\si{mm~mrad}] & $\Delta^{cum} \epsilon^y_N$ [\si{mm~mrad}] & $\Delta^{cum} \sigma_{\frac{\Delta E}{E}}$ [\si{\percent}]\\
        Beamline & Beam energy& $\Delta E$ & $\Delta^\text{cum} \epsilon^x_N$ & $\Delta^\text{cum} \epsilon^y_N$  & $\Delta^\text{cum} \sigma_{\frac{\Delta E}{E}}$ \\
       & [\si{GeV}] & [\si{MeV}]& [\si{mm~mrad}] & [\si{mm~mrad}] & [\%]\\
    \botrule  
    Arc 1 &  8.62 & 3 & 0.2 & 0.1 & 0.01\\
    Arc 2 & 16.73 & 25 & 1.0 & 1.2 & 0.03\\
    Arc 3 & 24.85 & 80 & 7.3 & 2.9 & 0.06\\
    Arc 4 & 32.96 & 229 & 13.2 & 14.0 & 0.12\\
   Arc 5 & 41.08 & 383 & 16.2 & 14.8 & 0.16\\
 \textbf{IR} & 49.19 & 39 & \textbf{16.2} & \textbf{18.7} & \textbf{0.18}\\
    Arc 6 & 49.19 & 797 & 25.2 & 22.6 & 0.24\\
    Arc 5 & 41.08 & 383 & 28.2 & 23.4 & 0.28\\
    Arc 4 & 32.96 & 229 & 34.1 & 34.5 & 0.33\\
    Arc 3 & 24.85 & 80 & 40.5 & 36.3 & 0.37\\
    Arc 2 & 16.73 & 25 & 41.2 & 37.4 & 0.39\\
    Arc 1 &  8.62 & 3 & 41.4 & 37.4 & 0.40\\
    Dump  &  0.5  & & 41.4 & 37.4 & 0.40\\
\botrule
\end{tabular}
  \caption{Energy loss and cumulative emittance dilution (transverse and longitudinal) due to synchroton radiation at the end of a given beam-line (complete Arc including: $180^0$ arc, Spreader, Recombiner and Doglegs in arcs 1-4). The table covers the entire ER cycle: 3 passes 'up' + 3 passes 'down'. Cumulative emittance dilution values just before the IP (past Arc 5 and Spr 6), which are critical for the luminosity consideration are highlighted in 'bold'. That row accounts for contributions from Spr 6 (the last bending section before the IR) to energy loss, as well as the vertical and longitudinal emittance dilutions. Here, $\Delta \sigma_{\frac{\Delta E}{E}} = \sqrt{\frac{\Delta \epsilon_E^2}{E^2}}$}.
  \label{tab:CummEmittance}
\end{table}

Tab.~\ref{tab:CummEmittance} shows, the LHeC luminosity requirement of total transverse emittance dilution in either plane (normalized) at the IP (at the end of Arc 5), not to exceed \SI{20}~{mm~mrad} (hor: \SI{16.2}~{mm~mrad} and ver: \SI{18.7}~{mm~mrad}) is met by-design, employing presented low emittance lattices in both the arcs and switch-yards. 
In the case of the optimised one-step spreader design, another reduction - mainly of the vertical emittance budget -  is obtained, providing a comfortable safety margin of the design.

Finally, one can see from Eqs.~\eqref{eq:Emit_dil_2} and~\eqref{eq:Emit_dil_3} an underlying universal scaling of the transverse (unnormalized) and longitudinal emittance dilution with energy and arc radius; they are both proportional to $\gamma^5/\rho^2$.
This in turn, has a profound impact on arc size scalability with energy; namely the arc radius should scale as $\gamma^{5/2}$ in order to preserve both the transverse and longitudinal emittance dilutions, which is a figure of merit for a synchrotron radiation dominated ERL.

\subsubsection{Compensation of Synchrotron Radiation Losses}

\begin{figure}[th]
  \centering
  \includegraphics[width=0.85\textwidth]{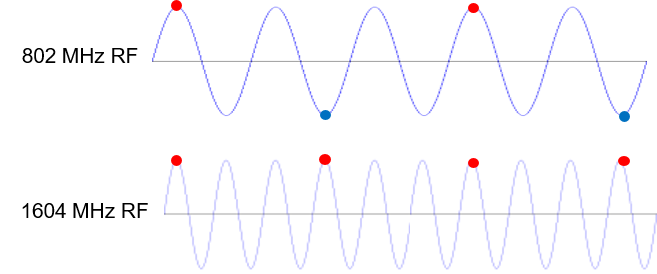}
  \caption{The second-harmonic RF restores the energy loss in both
    the accelerating and decelerating passes.}
  \label{fig:2_Harm}
\end{figure}

Depending on energy, each arc exhibits fractional energy loss  due to the synchrotron radiation, which scales as $\gamma^4/\rho$ (see Eq.~\eqref{eq:Emit_dil_1}).
Arc-by-arc energy loss was previously summarised in Tab.~\ref{tab:CummEmittance}. 
That energy loss has to be replenished back to the beam, so that at the entrance of each arc the accelerated and decelerated beams have the same energy, unless separate arcs are used for the accelerated and decelerated beams.
Before or after each arc, a matching section adjusts the optics from and to the linac.
Adjacent to these, additional cells are placed, hosting the RF compensating sections.
The compensation makes use of a second harmonic RF at \SI{1603.2}~~{MHz} to replenish the energy loss for both the accelerated and the decelerated beams, therefore allowing them to have the same energy at the entrance of each arc, as shown in Fig.~\ref{fig:2_Harm}.

Parameters of the RF compensation cryomodules,  
shown in Table ~\ref{tab:Harm__cryos}, have been extrapolated from the ILC cavity design, expecting that the higher frequency and lower gradient would support continuous operation. 
\begin{table}[!ht]
  \centering
%  \small
  \begin{tabular}{lcc} 
    \toprule
    Parameter & Unit & Value \\
    \botrule  
    Frequency & MHz & \SI{1603.2}{}\\
    Gradient & MV/m & \SI{30}{}\\
    Design &  & Nine cells\\
    Cells length  & mm & \SI{841}{}\\
    Structure length & m &  \SI{1}{}\\
    Cavity per cryomodule &  & {6}\\
    Cryomodule length & m & \SI{6}{}\\
    Cryomodule voltage & MV &\SI{150}{}\\
    \botrule  
  \end{tabular}
  \caption{ A tentative list of parameter for the compensating RF cryomodules extrapolated from the ILC design.}
  \label{tab:Harm__cryos}
\end{table}

As illustrated schematically in  Fig.~\ref{fig:2_Harm}, there are two beams in each arcs (with exception of Arc 6) one needs to replenish energy loss for: the accelerated and the decelerated beams. Assuming nominal beam current of \SI{20}{\milli\ampere}, the net current for two beams doubles. Therefore, \SI{40}{\milli\ampere} current in Arcs 1--5, was used to evaluated power required to compensate energy loss by 2-nd harmonic RF system, as summarized in Table~\ref{tab:Comp_cryos}.

\begin{table}[!ht]
  \centering
%  \small
  \begin{tabular}{lccc} 
    \toprule
    Section &  $\Delta E$ [\si{MeV}] & $P$ [\si{MW}] & Cryomodules \\
    \botrule  
    Arc 1 & 3 & 0.12 & 0\\
    Arc 2 & 25 & 1.0 & 0\\
    Arc 3 & 80 & 3.2 & 1\\
    Arc 4 & 229 & 9.16 & 2\\
    Arc 5 &  383 & 15.32 & 3\\
    Arc 6 & 836 & 16.7 & 6\\
    \botrule  
  \end{tabular}
  \caption{Arc-by-arc synchrotron radiated power for both the accelerated and decelerated beams (only one beam in Arc 6) along with a number of 2-nd harmonic RF cryomodules required to compensate energy loss.}
  \label{tab:Comp_cryos}
\end{table}
%\begin{table}[!ht]

The compensating cryomodules are placed into Linac 1 side of the racetrack, before the bending section of Arc~1, Arc~3, and Arc~5 and after the bending section of Arc~2, Arc~4, and Arc~6.
This saves space on Linac 2 side to better fit the IP line and the bypasses.
Note that with the current vertical separation of 0.5\,m it will not be possible to stack the cryomodules on top of each other; therefore, they will occupy 36\,m on the Arc~4 and Arc~6 side and 18 m on the Arc~3 and Arc~5 side of the racetrack.
Each of the compensating cavities in Arc~5 needs to transfer up to \SI{1}~~{MW} to the beam.
Although a \SI{1}~~{MW} continuous wave klystron are available~\cite{Zaltsman}, the cryomodule integration and protection system will require a careful design.
Tab.~\ref{tab:Comp_cryos} shows the energy loss for each arc and the corresponding synchrotron radiated power, along with number of cryomodules at \SI{1603.2}~~{MHz} RF frequency required to replenish the energy loss. 

\subsubsection{Component Summary}
%% ----
This closing section will summarise active accelerator components: magnets (bends and quads) and RF cavities for the \SI{50}~{GeV} baseline ERL. The bends (both horizontal and vertical) are captured in Tab.~\ref{tab:DipolesComponents}, while the quadrupole magnets and RF cavities are collected in Tab.~\ref{tab:QuadRFComponents}.

One would like to use a combined aperture (3-in-one) arc magnet design with 50\,cm vertical separation between the three apertures, proposed by Attilo Milanese~\cite{AM14}. That would reduce net arc bend count from 2112 to 704. As far as the Spr/Rec vertical bends are concerned, the design was optimised to include an additional common bend separating the two highest passes. So, there are a total of 8 trapezoid B-com magnets, with second face tilted by $3^0$ and large \SI{10}~{cm} vertical aperture, a, the rest are simple rectangular bends with specs from the summary Tab.~\ref{tab:DipolesComponents}.

\begin{table}[!ht]
  \centering
  \small
  \begin{tabular}{lccccccccc} 
    \toprule
    & \multicolumn{4}{c}{Arc dipoles} & & \multicolumn{4}{c}{Spr/Rec dipoles} \\
     \toprule
    Section & $N$ & $B [\text{T}]$ & $g/2[\text{cm}]$ & $L [\text{m}]$ &
    & $N$ & $B [\text{T}]$ & $g/2[\text{cm}]$ & $L [\text{m}]$ \\
    \botrule
    Arc 1 & 352 & 0.087 & 1.5 & 3 & & 8 & 0.678 & 2 & 3 \\
    Arc 2 & 352 & 0.174 & 1.5 & 3 & & 8 & 0.989 & 2 & 3 \\
    Arc 3 & 352 & 0.261 & 1.5 & 3 & & 6 & 1.222 & 2 & 3 \\
    Arc 4 & 352 & 0.348 & 1.5 & 3 & & 6 & 1.633 & 2 & 3 \\
    Arc 5 & 352 & 0.435 & 1.5 & 3 & & 4 & 1.022 & 2 & 3 \\
    Arc 6 & 352 & 0.522 & 1.5 & 3 & & 4 & 1.389 & 2 & 3 \\
    \botrule
    Total & 2112 &  &  &  &  & 36 &  &   \\
    \botrule
  \end{tabular}
  \caption{\SI{50}{GeV} ERL -- Dipole magnet count along with basic magnet parameters: Magnetic field $(B)$, Half-Gap $(g/2)$, and Magnetic length $(L)$.}
  \label{tab:DipolesComponents}
\end{table}

\begin{table}[!ht]
\centering
  \small
  \begin{tabular}{lccccccccc} 
    \toprule
    & \multicolumn{4}{c}{Quadrupoles} &  & \multicolumn{4}{c}{RF cavities} \\
    \botrule
    Section & $N$ & $G [\text{T/m}]$ & $a[\text{cm}]$ & $L [\text{m}]$ & &  $N$ & $f [\text{MHz}]$ & cell & $G_\text{RF}[\text{T/m}]$ \\
    \toprule
    Linac 1 &  29 &  7.7 & 3   & 0.25 & & 448 & 802 & 5 & 20 \\
    Linac 2 &  29 &  7.7 & 3   & 0.25 & & 448 & 802 & 5 & 20 \\
    Arc 1   & 255 &  9.25 & 2.5 & 1 & &   &   &   &   \\
    Arc 2   & 255 & 17.67 & 2.5 & 1 & &  &  &  &  \\
    Arc 3   & 255 & 24.25 & 2.5 & 1 & & 6 & 1604 & 9 & 30 \\
    Arc 4   & 255 & 27.17 & 2.5 & 1 & & 12 & 1604 & 9 & 30 \\
    Arc 5   & 249 & 33.92 & 2.5 & 1 & & 18 & 1604 & 9 & 30 \\
    Arc 6   & 249 & 40.75 & 2.5 & 1 & & 36 & 1604 & 9 & 30 \\
     \botrule
    Total & 1576 &  &  &  & & 968 &  &  & \\
    \botrule
  \end{tabular}
  \caption{\SI{50}{GeV} ERL -- Quadrupole magnet and RF cavities count along with basic magnet/RF parameters: Magnetic field gradient $(G)$, Aperture radius $(a)$, Magnetic length $(L)$, Frequency $(f)$, Number of cells in RF cavity (cell), and RF Gradient $(G_\text{RF})$.}
  \label{tab:QuadRFComponents}
\end{table}

\subsection{Interaction Region}

%  ---------------------------------
The Interaction Region (IR) of the ERL is one of the most challenging parts of the machine: While seeking for highest luminosity in ep-collisions, which includes strong mini-beta structures for both beams, the colliding bunches have to be separated and guided to their lattice structures, to avoid parasitic bunch encounters. In addition, collisions and beam-beam effects with the second non-colliding proton beam have to be avoided.
In order to meet these requirements, the design of the IR has been based on a compact magnet structure of focusing and bending fields which are optimized for an effective beam separation and smallest synchrotron radiation power and critical energy at the Interaction Point (IP).
Following the design of the LHC upgrade project, HL-LHC, and the conditions set by the technical feasibility of the beam separation scheme, the parameter list of the LHeC has been defined, Tab.~\ref{tab:design_param},  leading to a luminosity at the e-p interaction point in the order of $L= 10^{34} \si{ cm^{-2} s^{-1}}$.

%-----------------
\begin{table}[!h]
    \centering
    \begin{tabular}{c|c|c|c}
    \hline
    Parameter               & Unit             & Electrons & Protons \\
    \hline
    beam energy             & GeV              & 50        & 7000    \\ 
    beam current            &  mA              & 20        & 1400    \\
    bunches per beam        & -                & 1188      & 2808    \\
    bunch population        & $10^{10}$        & 0.3       &  22     \\
    bunch charge            & nC               & 0.5       &  35.24  \\
    norm. emittance (at IP) & mm $\cdot$ mrad  & 30        & 2.5     \\
    beta function at IP     & cm               &  10.9     &  10     \\
    beam-beam disruption    & -                & 14.3      & $1 \cdot 10^{-5}$ \\
    \hline
    luminosity              & $cm^{-2} s^{-1}$ &\multicolumn{2}{c}{$0.7\cdot10^{34}$} \\
    \hline
    \end{tabular}
    \caption{Parameter list of the LHeC}
    \label{tab:design_param}
\end{table}
% ----------------

\subsubsection  {Electron Beam Optics and Separation Scheme} 
\begin{figure}[tbh]
  \centering
  \includegraphics[width=0.99\textwidth]{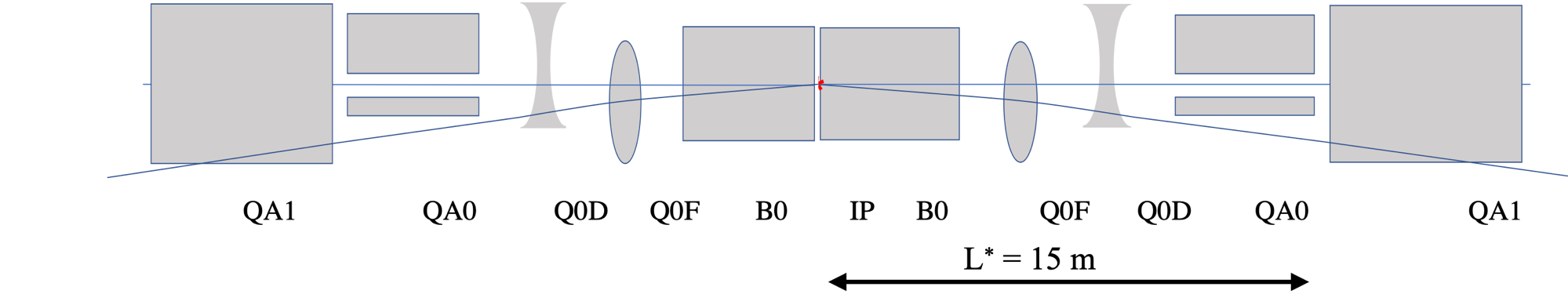}
  \caption{Schematic view of the combined focusing - beam separation scheme}
  \label{fig:IR_schematic}
\end{figure}

The design of the IR has to take a manifold of conditions into account: Focus the electron beam to the required $\beta$ values in both planes, establish sufficient beam separation, optimise the beam separation for smallest critical energy and synchrotron light power, and leave sufficient space for the detector hardware. A separation scheme has been established ~\cite{Thesis_Andre} that combines these requirements in one lattice structure (Fig.~\ref{fig:IR_schematic}). Due to the different rigidity of the beams, a separation is possible by applying a series of magnets, acting as a quasi-constant deflecting field: The spectrometer dipole of the LHeC detector, named {\it B0} in the figure, is used to establish a first separation. Right after and as close as possible to the IP, the mini-beta quadrupoles of the electron beam are located. They provide focusing in both planes for matched beam sizes of protons and electrons at the IP, $\beta_x(p) =\beta_x(e),\beta_y(p) =\beta_y(e)$. On top of that they are positioned off-center with respect to the electron beam, thus acting as combined function magnets to provide the same bending radius as the separator dipole: A continuous soft bending of the electron beam is achieved throughout the complete magnet structure. Additional conditions were put for a reduced beam size of the electron beam at the location of the first proton quadrupole. At this position, $L^{*} = \si{15} {m}$, the reduced electron beam size leads automatically to a minimum of the required beam separation and as direct consequence to smallest synchrotron radiation effects. 

The synchrotron light parameters, i.e. critical energy, radiation power and the geometry of the emitted light cone were determined with the simulation code BDSIM ~\cite{BDSIM}. As expected, the synchrotron light conditions in the arcs of the ERL become more serious turn by turn, reaching the highest level in the return arc after the collision point, where the \si{50}{GeV} electron beam is bent back to the first decelerating passage of the energy recovery process. The values of the emitted light are summarised in Tab.~\ref{tab:syli_param} and show the advantage of the ERL concept compared to traditional storage ring designs. As the final energy of the electron beam is reached at the very last turn only, and as the emitted light power is proportional to the beam energy $\gamma^4$, high synchrotron light losses essentially occur in a small part of the machine, namely the last return arc only.  

\begin{table}[!h]
    \centering
    \begin{tabular}{c|c|c|c}
    \hline
    Arc               & beam energy & crit. photon energy & Power of emitted light \\
                 &        (GeV)              &         (keV)          & (MW)                       \\ 
    \hline
    1            &         8.75              &           3.2          & 0.01     \\ 
    2            &         17                &           23.9         & 0.21     \\
    3            &         25.25             &           78.5         & 0.75     \\
    4            &         33.5              &           183.30       & 2.45     \\
    5            &         41.75             &           354.8        & 5.87     \\
    6            &         50                &           609.3        & 12.17    \\
    \hline
    \end{tabular}
    \caption{Critical energy and power of the emitted synchrotron light in the return arcs of the ERL.}
    \label{tab:syli_param}
\end{table}

\begin{figure}[tbh]
  \centering
  \includegraphics[width=0.65\textwidth]{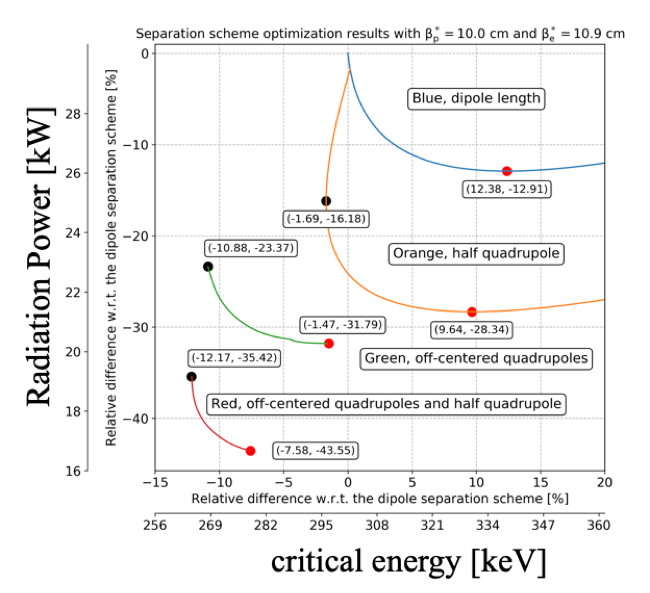}
  \caption{Optimising the synchrotron light for lowest critical energy and power, details in the text.}
  \label{fig:Sy-Li}
\end{figure}

Special care however is needed in the vicinity of the particle detector. The properties of the focusing elements, the separation scheme and the geometry of the Interaction Region (IR) have been optimised for smallest critical energies and power. Fig.~\ref{fig:Sy-Li} summarizes the results: The graph shows the reduction of the critical energy and power due to the different steps of the optimisation procedure. Starting from a pure separator dipole design to establish the required beam separation, the concept of a half-quadrupole as first focusing element in the proton lattice is introduced as well as an improved beam separation of the electrons by off-centre quadrupoles. The actual distribution of the detector dipole field and the off-centre quadrupoles has a considerable effect: The red and black points in the graph correspond to the minimum achievable critical energy and emitted power, respectively. Dedicated calculation of the synchrotron light cone and a sophisticated machine detector interface including absorbers have been performed to shield the detector parts and accelerator modules.

\subsubsection {Proton Beam Optics} 
\begin{figure}[tbh]
  \centering
  \includegraphics[width=0.65\textwidth]{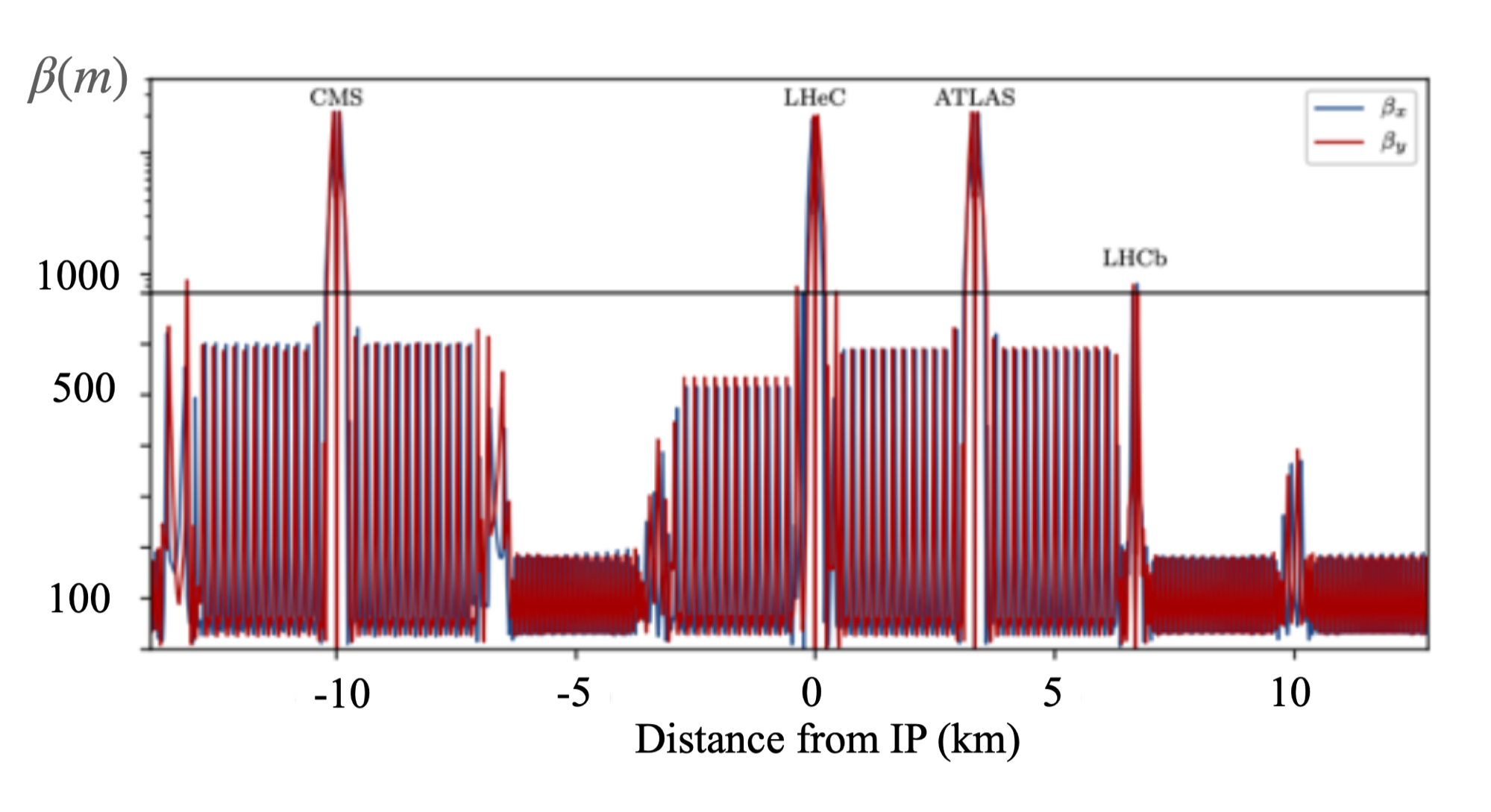}
  \caption{LHC proton beam optics, optimised for the LHeC design values of $\beta$=7 cm at the LHeC IP.}
  \label{fig:ats}
\end{figure}

The optics of the colliding proton beam follows the standard settings of the HL-LHC and is based on the so-called ATS scheme (achromatic telescoping squeeze), where additional focusing strength - and thus smaller vaues of $\beta^*$ at a given collision point - are obtained using the matching quadrupoles of the neighboring LHC octants. Fig.~\ref{fig:ats} shows the proton optics for values of e.g. $\beta^*=\si{7}{cm}$ at the interaction point of the LHeC - embedded and well matched into the HL-LHC optics for the ATLAS and CMS interaction points. The long-ranging beta-beat which is an essential feature of the HL-LHC optics ~\cite{HL-LHC} is clearly visible on both sides of the IP.

The operation of the LHeC electron-proton collisions is foreseen in parallel running mode to the LHC standard p-p operation. As a consequence, the design orbit of the second ``non-colliding'' proton beam at the e-p interaction point has to be included in the e-p IR layout. The basic principle of the LHC design remains unchanged and in order to preserve the overall geometry, the two proton beams are brought onto intersecting orbits at each IP, by strong separation and recombination dipoles. At the e-p interaction region, a collision of the two proton beams is avoided by selecting appropriately its location: Shifted in position and thus in time, direct collisions between the two proton beams as well as with the electron beam can be excluded. Long range encounters are suppressed by a large crossing angle of 7 mrad. The large crossing angle keeps the long range beam-beam effect small and separates the beams enough to allow septum quadrupoles to focus only the colliding beam. All in all the new e-p interaction region, including the mini-beta structure of the electron beam, is embedded in the existing LHC lattice to allow for concurrent e-p and p-p collisions
in the LHC interaction points.

\subsubsection {Beam-Beam Effects} 
\begin{figure}[tbh]
  \centering
  \includegraphics[width=0.65\textwidth]{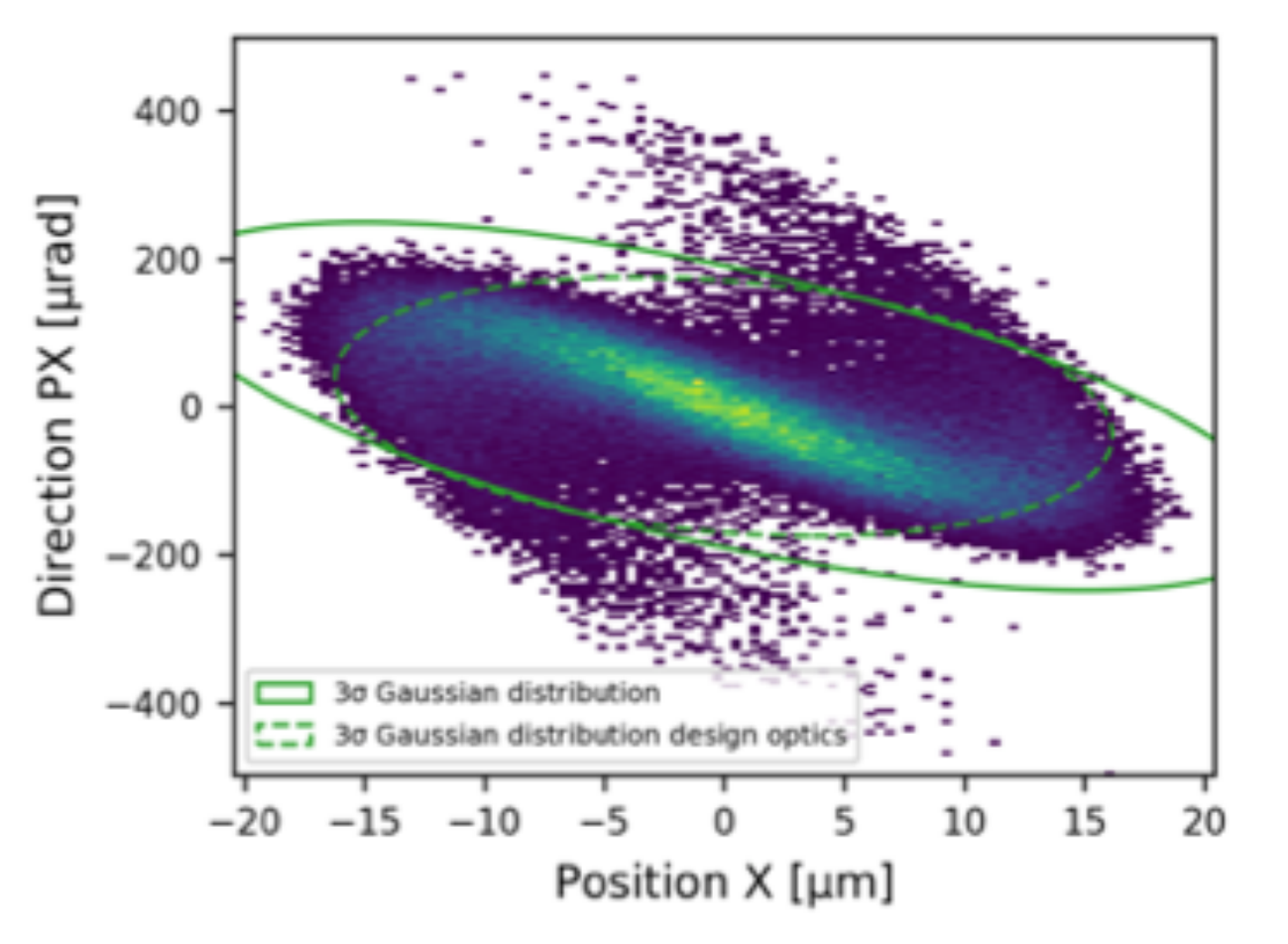}
  \caption{Simulation of the beam-beam effect for the electrons plotted in phase space coordinates, x,x’.}
  \label{fig:beam-beam}
\end{figure}

The beam-beam effect will always be the final limitation of a particle collider and care has to be taken, to preserve the beam quality of both, proton and electron beam. As concurrent operation of e-p collisions is foreseen in parallel to the LHC standard proton-proton operation, the beam beam effect of the protons has to be limited to preserve the proton beam emittance and allow successful data taking in the p-p collision points. Due to the limited bunch population of the electron beam, this is fulfilled by design.  
In the case of the electron beam the beam-beam effect is determined by the proton bunch population, which is considerably higher than the electron bunch intensity and  its detrimental effects on the electron emittance had to be limited to assure a successful energy recovery process in the ERL. In order to minimise the so-called beam disruption effect, the optical functions at the IP have been optimised, taken into account the influence of the beam-beam force. In Fig.~\ref{fig:beam-beam} the situation is represented in the (x,~x’) phase space. While tails in the transverse beam distribution as consequence of the beam-beam effect are clearly visible, the core of the beam still remains in a quasi ellipse like boundary. The coordinates obtained are used as starting conditions for the deceleration part of the ERL for a full front-to-end simulation. Given the design parameters of the LHeC, summarised in  Tab.~\ref{tab:design_param}, up to 99 \% of transmission efficiency - and thus an equivalent high value for the energy recovery process - have been achieved.  

\subsection{Civil Engineering}

\subsubsection{Introduction}

Since the beginning of the LHeC concept, various shapes and sizes of the eh collider were studied around CERN region. The conceptual study report published in 2012 focused primarily on two main options, namely the RING-RING and the RING-LINAC options. For civil engineering, these options were studied taking into account geology, construction risks, land features as well as technical constrains and operation of the LHC. The Linac-Ring conﬁguration was selected as preferred due to higher achievable luminosity (see Chapter 1.1). 

\begin{figure}[!ht]
  \centering
  \includegraphics[width=0.85\linewidth]{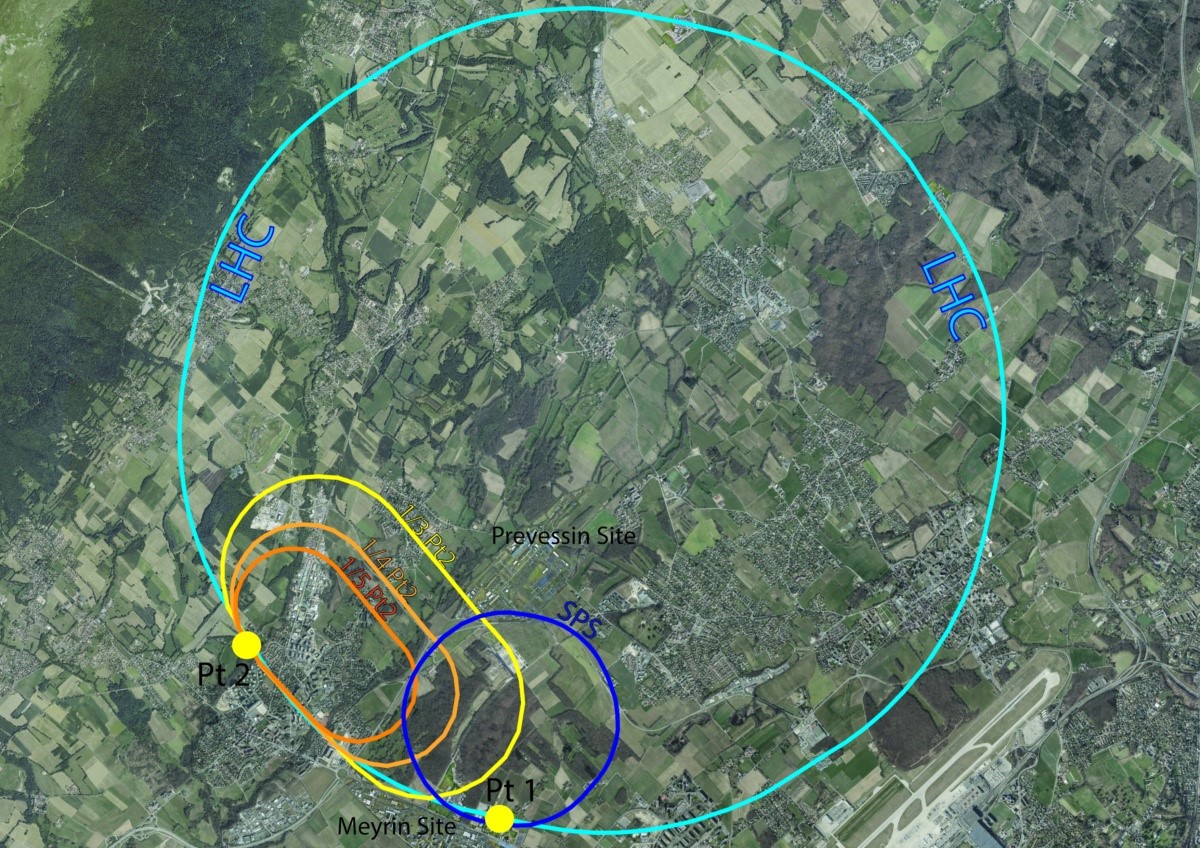}
  \caption{Three racetrack alternatives proposed for the eh machine at LHC Point 2}
   \label{fig:Racetrack_alt}
\end{figure}

This chapter describes the civil engineering infrastructure required for an Energy Recovery Linac (ERL) injecting into the LHC ALICE cavern at LHC Point 2. Figure~\ref{fig:Racetrack_alt} shows three options of different sizes proposed for the ERL, represented as fractions of the LHC circumference. This chapter focuses on two of these options, speciﬁcally the 1/3 and 1/5 of the LHC circumference. 

 \subsubsection{Placement and Geology }
 
 The proposed siting for the LHeC is in the North-Western part of the Geneva region at the existing CERN laboratory. The proposed Interaction Region is fully located within existing CERN land at LHC Point 2, close to the village of St. Genis, in France. The CERN area is extremely well suited to housing such a large project, with well understood ground conditions having several particle accelerators in the region for over 50 years. Extensive geological records exist from previous projects such as LEP and LHC and more recently, further ground investigations have been undertaken for the High-Luminosity LHC project. Any new underground structures will be constructed in the stable molasse rock at a depth of 100-150m in an area with very low seismic activity.
 
\begin{figure}[!ht]
    \centering
    \includegraphics[width=0.9\linewidth]{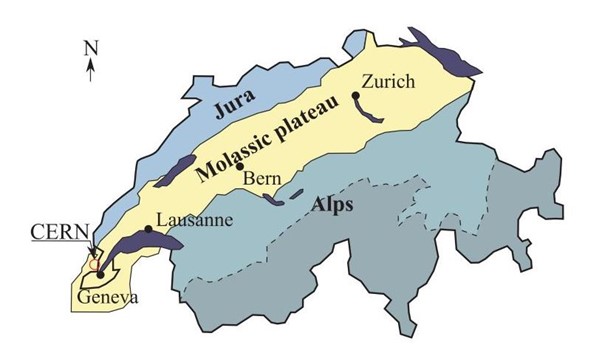}
    \caption{Schematic description of Swiss geology}
   \label{fig:Swiss_geology}
\end{figure}

\begin{figure}[!ht]
    \centering
    \includegraphics[width=0.9\linewidth]{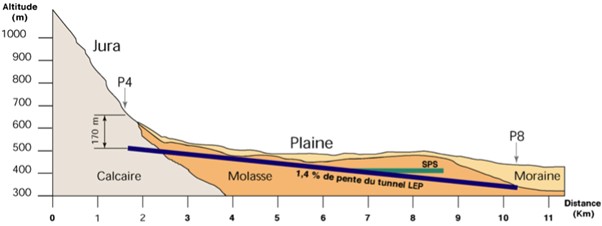}
    \caption{Schematic description of Swiss geology}
\label{fig:Swiss_geology_1}
\end{figure}

The LHeC is situated within the Geneva basin, a sub-basin of the large molassic plateau (Figure~\ref{fig:Swiss_geology}). The molasse formed from the erosion of the Alps and it is a weak sedimentary rock. It comprises of alternating layers of marls and sandstones (and formations of intermediate compositions), which show a high variety of strength parameters. The molasse is overlaid by the Quaternary glacial deposits called moraines. Figure~\ref{fig:Swiss_geology_1} shows a simpliﬁed geological proﬁle of the LHC. Although placed mainly within the molasse plateau, one sector of the LHC is situated in the limestone of the Jura.

The physical positioning of the LHeC has been developed based on the assumption that the maximum underground volume possible should be placed within the Molasse Rock and should avoid as much as possible any known geological faults or environmentally sensitive areas. Stable and dry, the molasse is considered a suitable rock type for TBM excavation.  In comparison, CERN has experienced significant issues with the underground construction of sector 3-4 in the Jura limestone. There were major issues with water ingress at and behind the tunnel face. Another challenging factor for limestone is the presence of karsts. They are the result of chemical weathering of the rock and often they are filled with water and sediment, which can lead to strong water inflows and instability of the excavation. 

\begin{figure}[!ht]
    \centering
    \includegraphics[width=0.9\linewidth]{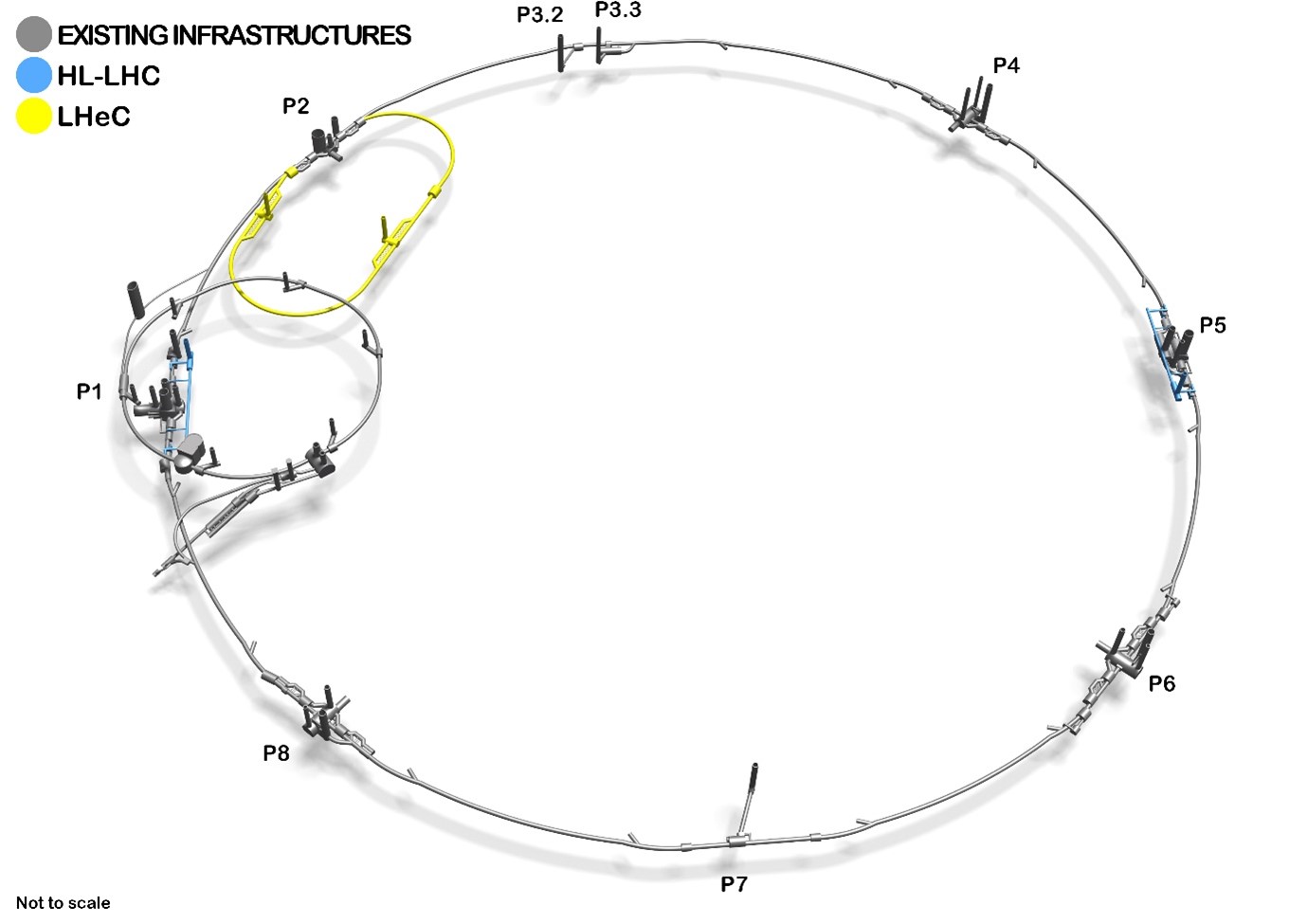}
    \caption{3D schematic showing the proposed racetrack of the Large Hadron electron Collider at high luminosity. }
   \label{fig:3D_LHC}
\end{figure}

The ERL will be positioned inside the LHC Ring, in order to ensure that new surface facilities are located on existing CERN land. The proposed underground structures for a Large Hadron electron Collider (LHeC) at high luminosity aiming for an electron beam energy of 60 GeV is shown in Figure~\ref{fig:3D_LHC}. The LHeC tunnel will be tilted similarly to the LHC at a slope of 1.4\% to follow a suitable layer of molasse rock.

 \subsubsection{Underground infrastructure}
The underground structures proposed for LHeC option 1/3 LHC  require a tunnel approximately 9~km long of 5.5~m diameter, including two LINACs.  Parallel to the main LINAC tunnels, at 10~m distance apart, there are the RF galleries, each 1070~m long. Waveguides of 1m diameter are connecting the RF galleries and LHeC main tunnel. These structures are listed in Table~\ref{tab:Table_12}.
Two additional caverns, 25~m wide and 50~m long are required for cryogenics and technical services. These are connected to the surface via two 9~m diameter access, provided with lifts to allow access for equipment and personnel. 

Additional caverns are needed to house injection facilities and a beam dump. The underground structures proposed for LHeC option 1/5 LHC are the same as 1/3 options with the exception of the main tunnel which would be 5.4~km long connected to RF galleries, each 830~m long.  

\begin{figure}[!ht]
    \centering
    \includegraphics[width=0.9\linewidth]{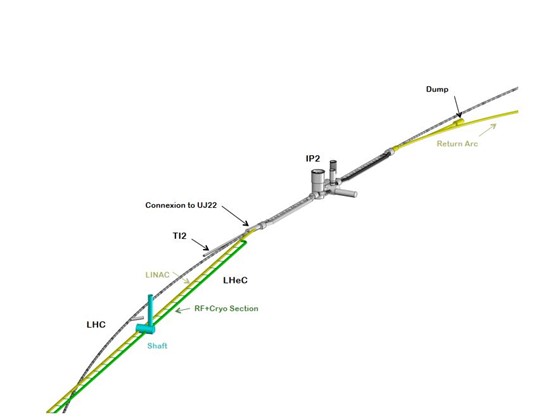}
    \caption{ERL injection area into IP2 and RF/Cryo/Services Cavern}
   \label{fig:ERL_injection}
\end{figure}

\begin{figure}[!ht]
    \centering
    \includegraphics[width=0.9\linewidth]{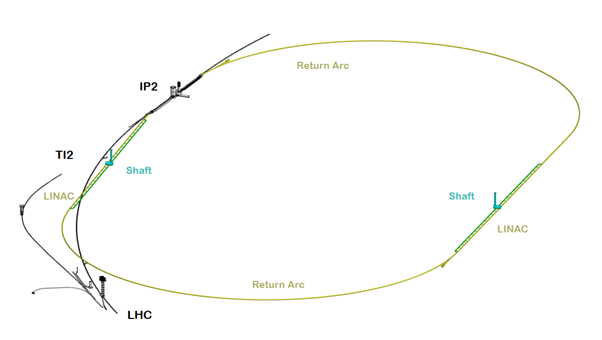}
    \caption{Drawing showing the underground structures for LHeC  (1/4 LHC option)}
   \label{fig:Underground_Structures}
\end{figure}

In addition to the new structures, the existing LHC infrastructure also requires modifications. To ensure connection between LHC and LHeC tunnels, the junction caverns UJ22 and UJ27 need to be enlarged (Figures~\ref{fig:ERL_injection} and \ref{fig:Underground_Structures}). Localised parts of the cavern and tunnel lining will be broken out to facilitate the excavation of the new spaces and the new connections, requiring temporary support.

Infrastructure works for LEP were completed in 1989, for which a design lifespan of 50 years was specified. If LHC is to be upgraded with a high energy, refurbishment, maintenance works are needed to re-use the existing infrastructure. 
Shaft locations were chosen such that the surface facilities are located on CERN land.  The scope for surface sites is still to be defined. New facilities are envisaged for housing technical services such as cooling and ventilation, cryogenics and electrical distribution. 

\begin{table}[!ht]
    \centering
    \begin{tabular}{c|c|c|c}
    \hline
    Structures & Quantities & Lenght & Span \\
    \hline
    Machine tunnels & - & 9091~m & 5.5~m ID\\ 
    \hline
    Service caverns & 2 & 50~m & 25~m\\
    \hline
    Service shafts & 2 & 80~m & 9~m ID\\
    \hline
    Injection cavern & 1 & 50~m &25~m\\
    \hline
    Dump cavern & 1 & 90~m & 16.8~m\\
    \hline
    RF Galleries & 2 & 1070~m & 5.5~m\\
    \hline
    Waveguide Connections & 50 & 10~m & 1~m ID\\
    \hline
    Connection galleries & 4 & 10~m & 3~m ID\\
    \hline
    Junction Caverns & 3 & 20~m & 16.8~m\\
    \hline
    \end{tabular}
    \caption{List of structures for 1/3 LHC option}
    \label{tab:Table_12}
\end{table}

\subsubsection{Construction Methods}
A Tunnel Boring Machines (TBM) should be utilised for the excavation of the main tunnel to achieve the fastest construction. When ground conditions are good and the geology is consistent, TBMs can be two to four times faster than conventional methods. A shielded TBM could be employed, with pre-cast segmental lining, and injection grouting behind the lining. 

\begin{figure}[!ht]
    \centering
    \includegraphics[width=0.8\linewidth]{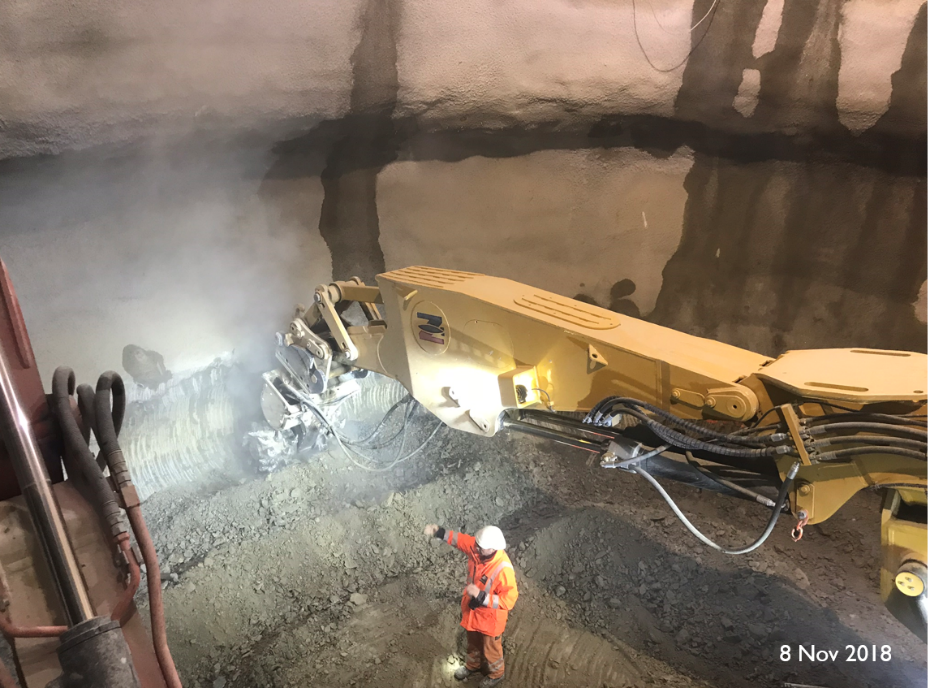}
    \caption{Excavator with hydraulic cutting heads being used at HL-LHC Point 1}
   \label{fig:Excavator_Hydraulic}
\end{figure}

\begin{figure}[!ht]
    \centering
    \includegraphics[width=0.8\linewidth]{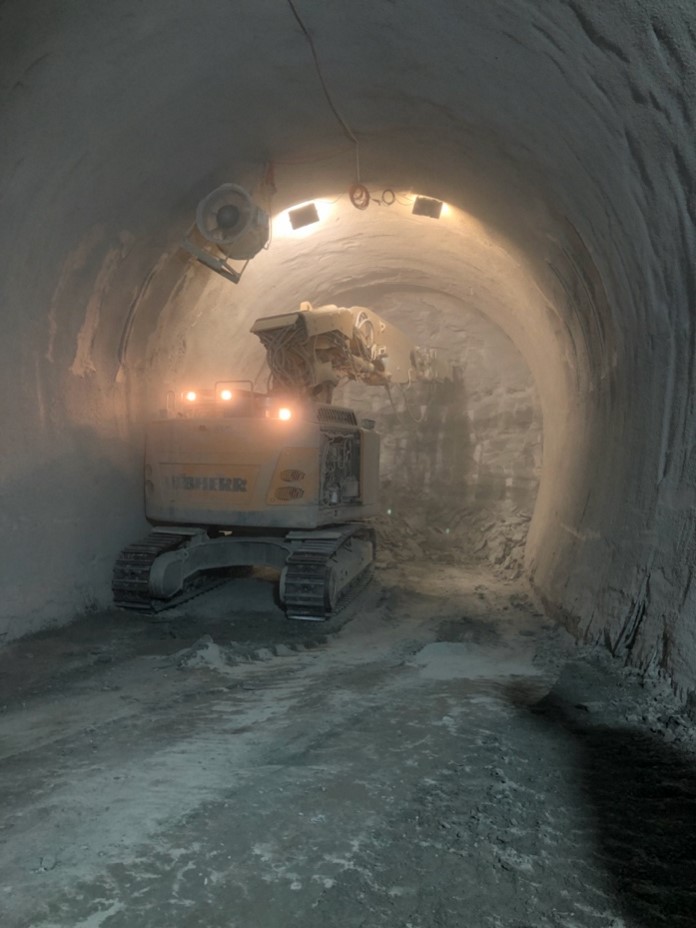}
    \caption{Rockbreaker used for cavern excavation at HL-LHC Point 1}
   \label{fig:Rockbreaker}
\end{figure}

For the excavation of the shafts, caverns and connection tunnels, conventional technique could be used. Similar construction methods as for HL-LHC, for example using roadheaders and rockbreakers, can be adopted for LHeC. Some of these machinery could be seen in Figures~   \ref{fig:Excavator_Hydraulic} and \ref{fig:Rockbreaker} showing the excavation works at point 1 HL-LHC. One main constraint that dictated what equipment to be used for the HL-LHC excavation, was the vibration limit.  Considering the sensitivity of the beamline, diesel excavators have been modified and equipped with an electric motor in order to reduce vibrations that could disrupt LHC operation. A similar equipment could also be needed for LHeC if construction works are carried out during operation of the LHC. 

Existing boreholes data around IP2 shows that the moraines layer can be 25-35~m deep before reaching the molasse. Temporary support of the excavation, for example using diaphragm walls is recommended. Once reaching a stable ground in dry conditions, common excavation methods can be adopted, for example using a roadheaders and rockbreakers.  The shaft lining will consist of a primary layer of shortcrete with rockbolts and an in-situ reinforced concrete secondary lining, with a waterproofing membrane in between the two linings. 

\subsubsection{Cost Estimate}
A cost estimate was prepared for a 9.1k~m ERL located at Point 2 of LHC, using the same measure prices as for FCC. More recently for LHeC, the cost figures were adapted to fit the smaller version, the 5.4~km racetrack at point 2 (option 1/5 ~LHC).  

The civil engineering costs amount to about 25\% of the total project costs. In particular, for a 9.1~km ERL (1/3 LHC option) the civil engineering was estimated at 386~MCHF and for a 5.4~km configuration (1/5~LHC) the costs are 289~MCHF. These estimates include the fees for preliminary design, approvals and tender documents (12\%), site investigations (2\%) and contractor’s profit (3\%). The costs mentioned do not include surface structures.  Where possible, existing surface infrastructure will be re-used. 

\section{PERLE -- A 'Stepping Stone' for the LHeC}
PERLE (Powerful ERL for Experiments) \cite{CDR:PERLE} is a novel Energy Recovery Linac (ERL) test facility, designed to validate choices for a 50~GeV ERL foreseen in the design of the Large Hadron Electron Collider (LHeC) and the Future Circular Collider (FCC-eh), and to host dedicated nuclear and particle physics experiments. Its main thrust is to probe high current, continuous wave (CW), multi-pass operation with superconducting cavities at 802~MHz. With very high transient beam power (10~MW), PERLE offers an opportunity for controllable study of every beam dynamic effect of interest in the next generation of ERL design and becomes a ‘stepping stone’ between present state-of-art 1~MW ERLs and future 100~MW scale applications. 
Particularly, the PERLE facility, to be hosted at Ir\`{e}ne Joliot Curie Laboratory, targets the LHeC configuration and beam currents of up to 20\,mA (corresponding to a 120\,mA cavity load).
This unique quality beam is intended to perform a number of experiments in different fields; ranging from uncharted tests of accelerator components via elastic ep scattering to laser-Compton back-scattering for photon physics \cite{Erk:Exp}. Following an experiment, the CW beam will be decelerated in three consecutive passes back to the injection energy, transferring virtually stored energy back to the RF.
\subsection{PERLE Facility}
\begin{figure}
    \centering
    \includegraphics[width=1.05\textwidth]{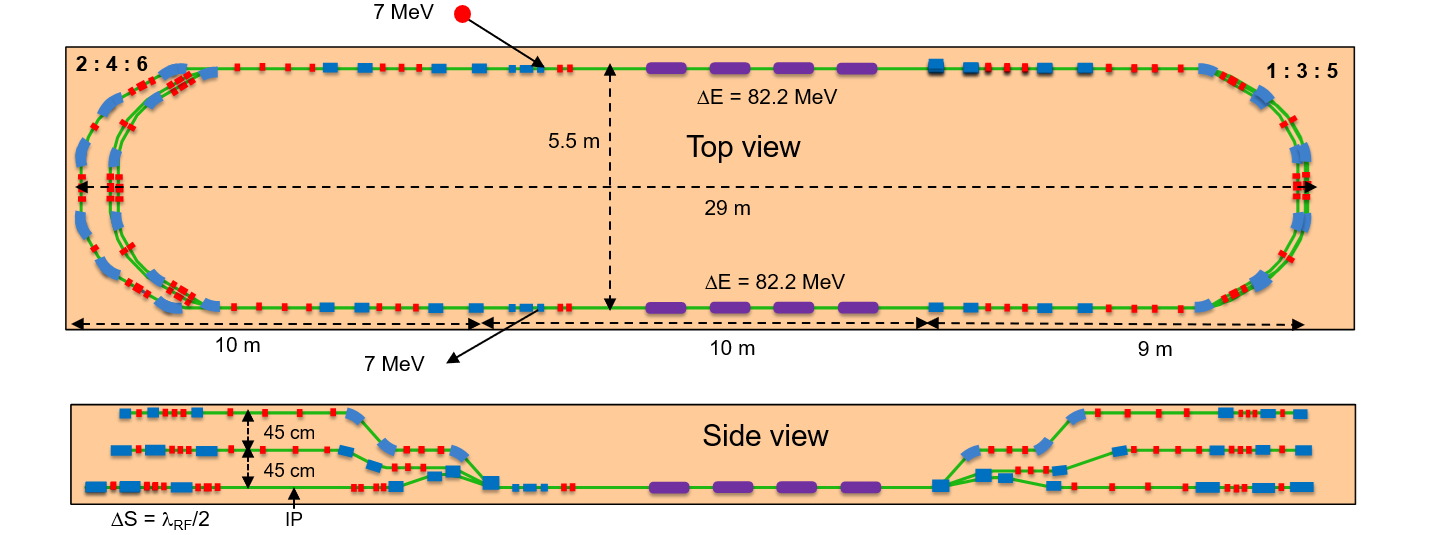}
    \caption{Top and side views of PERLE, featuring two parallel linacs each hosting a 82.2 MeV cryomodule, achieving 500 MeV in three passes.}
  \label{fig:PERLElayout}
\end{figure}

PERLE accelerator complex is arranged in a racetrack configuration; hosting two cryomodules (containing four, 5-cell, cavities operating at 802 MHz), each located in one of two parallel straights, completed with a stack of three recirculating arcs on each side (with 45 cm vertical separation between the arcs). Additional space is taken by 4-6 meter long spreaders/recombiners, including matching sections and two experimental areas, as illustrated in Fig.~\ref{fig:PERLElayout}. 

PERLE optics features Flexible Momentum Compaction (FMC) lattice architecture \cite{LHeC:Optics} for six vertically stacked return arcs. Starting with a high current (in excess of 20 mA) 7 MeV photo-injector, final energy of 500 MeV can be reached in three re-circulation passes, assuming a 4-cavity cryomodules. Each of the two cryomodules provides 82.2 MeV energy boost. 
A summary of design parameters is presented in Table~\ref{tab:parameters}. The beam parameters have been chosen to match those of LHeC \cite{CDR:2020}, so that it will serve as a test bed for the ERL design and SRF technology development. The bunch spacing in the ERL is assumed to be \SI{25}~~{ns}, however empty bunches might be required in the ERL for ion clearing gaps.

\begin{table}[!hbt]
   \centering
   \caption{PERLE Beam Parameters}
   \begin{tabular}{lcc}
    \toprule
       \textbf{Parameter} & Unit & Value \\
   \botrule
      Injection beam energy & MeV & 7\\
       Electron beam energy & MeV & 500 \\
       Norm. emittance $\gamma\varepsilon_{x,y}$ & mm mrad & 6 \\
       Average beam current & mAmp & 20 \\
       Bunch charge & pCoulomb & 500 \\  
       Bunch length & mm & 3 \\
       Bunch spacing & nsec & 24.95 \\
       RF frequency & MHz & 801.58 \\
       Duty factor & & CW \\
   \botrule
   \end{tabular}
   \label{tab:parameters}
\end{table}

\subsection{Multi-pass Linac Optics with Energy Recovery}

\begin{figure}[!tbh]
    \centering
    \includegraphics*[width=0.95\textwidth]{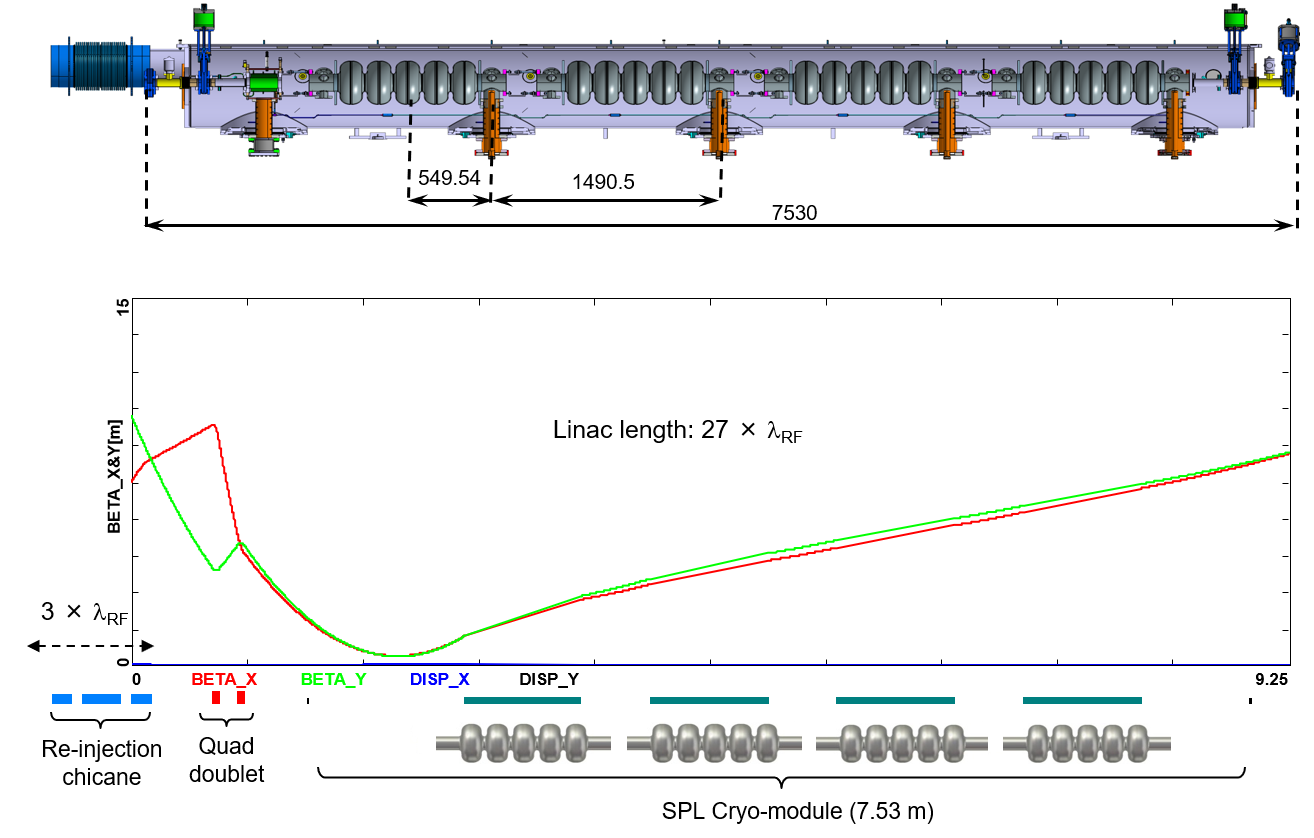}
    \caption{Linac configured with the SPL cryo-module. Injection, 1-st pass linac Optics tunable by an initial quadrupole doublet.}
  \label{fig:SPL_cryo_linac}
\end{figure}

Injection at \SI{7}~~{MeV} into the first linac is done through a fixed field injection chicane, with its last magnet (closing the chicane) being placed at the beginning of the linac. It closes the orbit ‘bump’ at the lowest energy, injection pass, but the magnet (physically located in the linac) will deflect the beam on all subsequent linac passes. In order to close the resulting higher pass ‘bumps’, the so-called re-injection chicane is instrumented, by placing two additional opposing bends in front of the last chicane magnet. This way, the re-injection chicane magnets are only ‘visible’ by the higher pass beams. Layout and injection pass optics is illustrated in Fig.~\ref{fig:SPL_cryo_linac} 
The second linac in the racetrack is configured exactly as a mirror image of the first one, with a replica of the re-injection chicane at its end, which facilitates a fixed-field extraction of energy recovered beam to the dump (at \SI{7}~~{MeV}). 

\begin{figure}[!tbh]
    \centering
    \includegraphics*[width=1.0\textwidth]{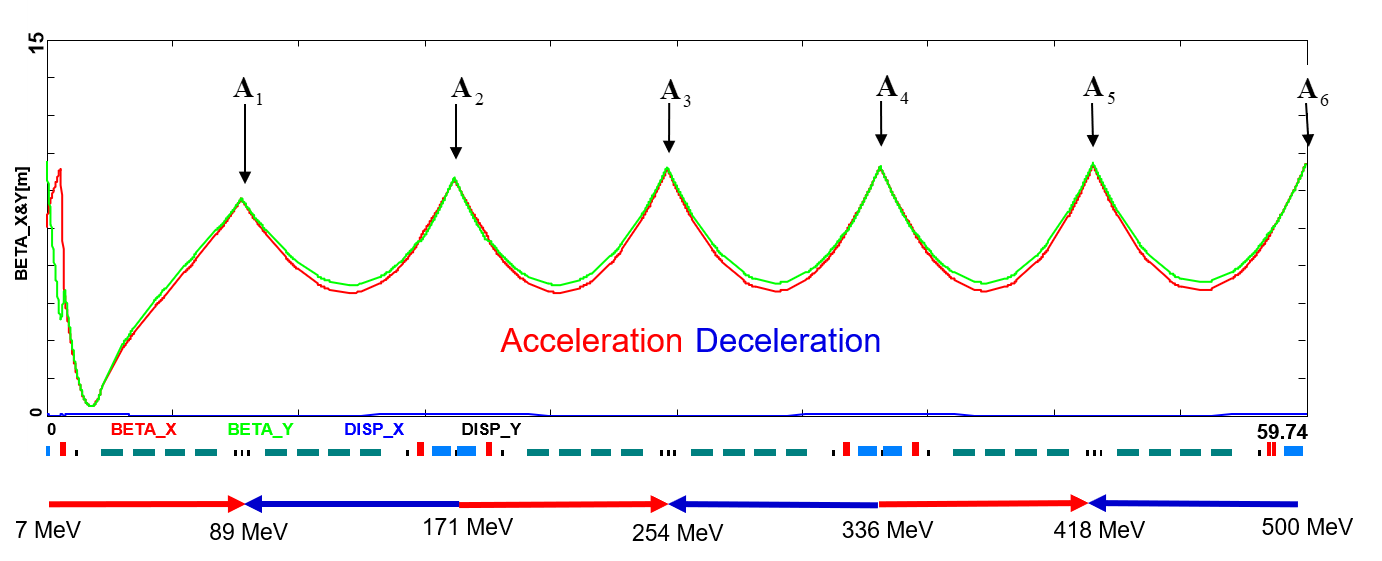}
    \caption{Multi-pass linac optics. Red/Green curves illustrate symmetrically optimized horizontal/vertical beta functions across different passes through the linac; Red/Blue arrows indicate the accelerating/decelerating passes.}
  \label{fig:Multi_pass_linac}
\end{figure}

 Multi-pass energy recovery in a racetrack topology explicitly requires that both the accelerating and the decelerating beams share the individual return arcs. This in turn, imposes specific requirements for the TWISS function at the linacs ends: the TWISS functions have to be identical for both the accelerating and decelerating linac passes converging to the same energy and therefore entering the same arc. 
To represent beta functions for multiple accelerating and decelerating passes through a given linac, it is convenient to reverse the linac direction for all decelerating passes and string them together with the interleaved accelerating passes, as illustrated in Fig.~\ref{fig:Multi_pass_linac}. This way, the corresponding accelerating and decelerating passes are joined together at the arcs entrance/exit, automatically satisfying the matching conditions into the arcs.

 \subsection{Recirculating Arc Architecture}
The spreaders are placed directly after each linac to separate beams of different energies and to route them to the corresponding arcs. The recombiners facilitate just the opposite: merging he beams of different energies into the same trajectory before entering the next linac.  Each spreader starts with a vertical bending magnet, common for all three beams, that initiates the separation.  The highest energy, at the bottom, is brought back to the initial linac level with a chicane.  The lower energies are captured with a two-step vertical beam line.  The vertical dispersion introduced by the first step bends is suppressed by the three quadrupoles located appropriately between the two steps. 

 \begin{figure}[!tbh]
    \centering
    \includegraphics*[width=1.0\textwidth]{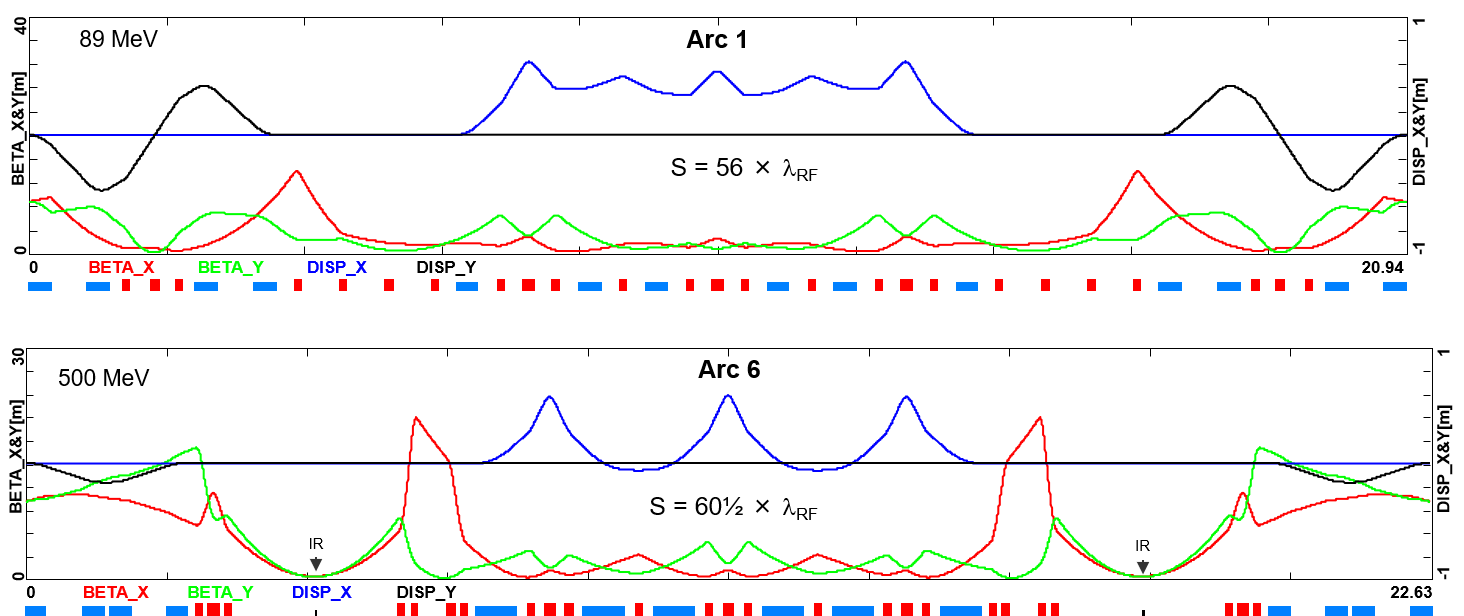}
    \caption{The lowest and highest energy arcs (arc 1 and arc 6). Optics architecture based on the FMC cell. Horizontal (red curve) and vertical (green curve) beta-function amplitudes are illustrated. Blue and black curves represent the horizontal and vertical dispersion, respectively. The arc, as configured above, is tuned to the isochronous condition, of zero momentum compaction factor.}
  \label{fig:Arcs_1_6}
\end{figure}

 The lowest energy spreader is configured with three curved bends following the common magnet, because of a large bending angle (30 deg.) the spreader is configured with. This minimizes adverse effects of strong edge focusing on dispersion suppression for a lower energy spreader. Following the spreader,there are four matching quads to ‘bridge’ the TWISS function between the spreader and the following 180 deg. arc (two betas and two alphas). 
All six, 180 deg. horizontal arcs are configured with a FMC style optics to ease individual adjustment of the momentum compaction factor in each arc (needed for the longitudinal phase-space re-shaping, essential for operation with energy recovery). 
The three arcs on either side of the linacs are vertically stacked and composed of 6 dipoles instead of 4 dipoles with respect to the previous design \cite{CDR:PERLE}. The increased number of dipoles allow to reduce the effects of CSR \cite{CSR:1}. The low energy implies that the energy spread and emittance growth due to incoherent synchrotron radiation is negligible in the arcs.

The lower energy arcs (1, 2, 3) are composed of six 33 cm long curved 30 deg. bends and of a series of quadrupoles (two triplets and one singlet), while the higher arcs (4, 5, 6) use ‘double length’, 66 cm long, curved bends.  The usage of curved bends is dictated by a large bending angle (30 deg.).  If rectangular bends were used, their edge focusing would have caused significant imbalance of focusing, which in turn, would have had adverse effect on the overall arc optics.  Another reason for using curved bends is to eliminate the problem of magnet sagitta, which would be especially significant for longer, 66 cm, bends.  Each arc is followed by a matching section and a recombiner (mirror symmetric to previously described spreader and matching section). As required in case of mirror symmetric linacs, matching conditions described in the previous section, impose a mirror symmetric arc optics (identical betas and sign reversed alphas at the arc ends).
Complete lattices for the lowest and highest energy arcs (arc 1 at \SI{89}{MeV} and arc 6 at \SI{500}{MeV}), including a spreader, 180 deg. horizontal arcs and a recombiner, are illustrated in Fig.~\ref{fig:Arcs_1_6}. Presented arc optics architecture features high degree of modular functionality to facilitate momentum compaction management, as well as orthogonal tunability for both the beta functions and dispersion. The path-length of each arc is chosen to be an integer number of RF wavelengths, except for the highest energy pass, arc 6, whose length is longer by half of the RF wavelength (to shift the RF phase from accelerating to decelerating, switching to the energy recovery mode). The optimal bunch recombination pattern gives some constraints on the length of the arcs.

\subsection{Experimental Areas}
\begin{figure}[!tbh]
    \centering
    \includegraphics*[width=1.0\textwidth]{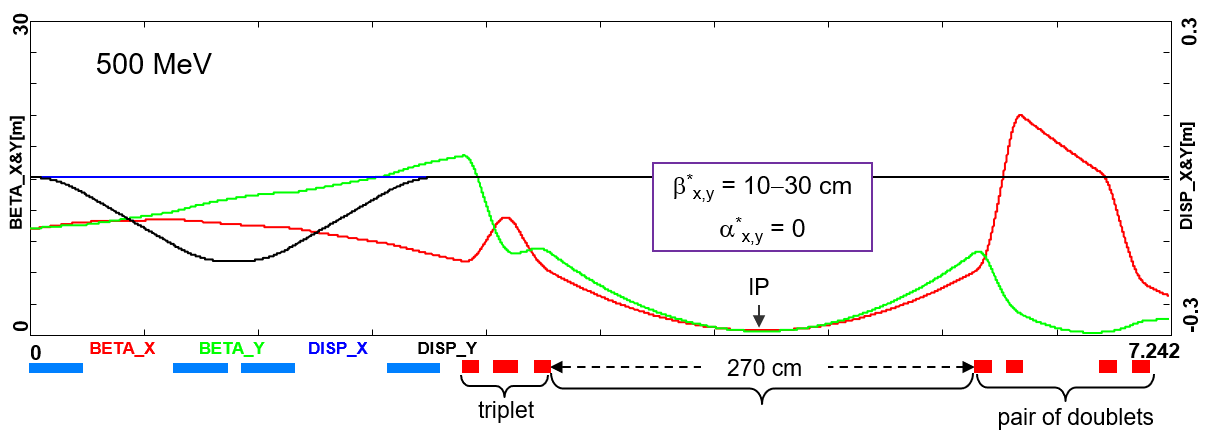}
    \caption{Optics design for 2.7 meter long Experimental Area, with a low-beta squeeze configured with a pair of doublets
}
  \label{fig:EA}
\end{figure}

PERLE facilitates a pair of Experimental Areas configured  at \SI{500}~{MeV}, located symmetrically on both sides of arc 6. Their optics based on low-beta squeeze configured with a pair of doublets, is illustrated in Fig.~\ref{fig:EA}.

\subsection{Outlook}
PERLE is a compact three-pass ERL test facility based on SRF technology, expanding the operational regime for multi-turn ERLs to around 10\,MW  of beam power. PERLE will serve as a hub for validation of a broad range of ERL accelerator phenomena, probing an unexplored operational regime and braking new grounds add developing novel ERL technology for future energy and intensity frontier machines. 
Innovative PERLE design expands on recently developed technological components, such as: 802 MHz Niobium cavity developed (JLab in collaboration with CERN) for the LHeC and FCC-ee, which features a high Q$_0$ of $3~10^{10}$ and an impressive gradient of nearly 30\,MV/m. The facility will initially use several in-kind contributions: a gun (from ALICE at Daresbury), a booster cryostat (from JLab) and a main linac cryostat (from CERN adapting the SPL module). 
The PERLE Collaboration has recently established an ambitious plan for first beam operation in the mid twenties. Several electron-scattering experiments are in the early phase of planning.
Integration of PERLE into the European Road-map for Accelerators is quite timely, since both the FCC-ee and recently the ILC are proposed as ERL colliders with significantly increased luminosity and substantially reduced power consumption. Needless to say, PERLE is positioned as a key effort towards future High Energy Physics, Particle Physics and Nuclear Physics facilities.

\section*{Funding Information}
Work at Jefferson Lab has been supported by the U.S. Department of
Energy, Office of Science, Office of Nuclear Physics under contracts
DE-AC05-06OR23177 and DE-SC0012704.
\section{Acknowledgments}
Useful discussions and conceptual input from:  Kevin André, Oliver Bruning, Max Klein, Walid Kaabi, Dario Pellegrini and Alessandra Valloni at all stages of the accelerator design process are greatly acknowledged.

\bibliographystyle{ws-rv-van}
\bibliography{ws-rv-sample}

%\blankpage
%\printindex[aindx]                 % to print author index
%\printindex                        % to print subject index
\end{document}